\documentclass[useAMS,usegraphicx,usenatbib]{mn2e}
\usepackage{rotating,graphicx,subfig,amsmath}

\catcode`\@=11
\def\gsim{\ifmmode{\mathrel{\mathpalette\@versim>}}
    \else{$\math
rel{\mathpalette\@versim>}$}\fi}
\def\lsim{\ifmmode{\mathrel{\mathpalette\@versim<}}
    \else{$\mathrel{\mathpalette\@versim<}$}\fi}
\def\@versim#1#2{\lower 2.9truept \vbox{\baselineskip 0pt \lineskip
    0.5truept \ialign{$\m@th#1\hfil##\hfil$\crcr#2\crcr\sim\crcr}}}
\catcode`\@=12

\defcitealias{st96}{ST96}
\defcitealias{dl07}{DL07}
\defcitealias{dl08}{DL08}
\defcitealias{dl09}{DL09}

\def\eg{{\it e.g.\ }}

\def\degrees{^\circ}

\def\eg{{\it e.g.\ }}

\def\ud{{\mathrm d}}

\newcommand{\greeksym}[1]{{\usefont{U}{psy}{m}{n}#1}}

\newcommand{\uDelta}{\mbox{\greeksym{D}}}

\title[Regularizing made-to-measure models]{Regularizing made-to-measure particle models of galaxies} 

\author[L. Morganti \& O. Gerhard]{Lucia Morganti$^{1}$\thanks{E-mail: morganti@mpe.mpg.de} and
Ortwin Gerhard$^1$\thanks{E-mail: gerhard@mpe.mpg.de}
\\$^1$ Max-Planck-Institut f\"ur extraterrestrische Physik, Postfach 1312, 
Giessenbachstr., D-85741 Garching, Germany }

\begin{document}   
   
\date{Accepted 2012 February 9.  Received 2012 February 9; in original form 2011 December 20.}

\pagerange{\pageref{firstpage}--\pageref{lastpage}} \pubyear{----}
  
\maketitle

\label{firstpage}

\begin{abstract}  

Made-to-measure methods such as the parallel code NMAGIC 
are powerful tools to build galaxy models reproducing observational data. 
They work by adapting the particle weights in an N-body system 
until the target observables are well matched. 
Here we introduce a moving prior regularization (MPR) method
for such particle models. 
It is based on determining from the particles a distribution of
priors in phase-space,
which are updated in parallel with the weight adaptation. This method allows
one to construct smooth models from noisy data 
without erasing global phase-space gradients. 
We first apply MPR to a spherical system for which
the distribution function can in theory be uniquely recovered
from idealized data. We show that NMAGIC with MPR indeed converges to
the true solution with very good accuracy, independent of the initial
particle model. Compared to the standard weight entropy
regularization, biases in the anisotropy structure are removed and
local fluctuations in the intrinsic distribution function are reduced. 
We then investigate how the uncertainties in the inferred
dynamical structure increase with less complete and noisier kinematic
data, and how the dependence on the initial particle model also increases.
Finally, we apply the MPR technique to the two
intermediate-luminosity elliptical galaxies NGC 4697 and NGC 3379,
obtaining smoother dynamical models in luminous and dark matter potentials.
\end{abstract}   


\section{Introduction}   
\label{sec:intro}   

In galactic dynamics, the modelling of photometric and kinematic observations 
is of great importance to infer intrinsic properties of galaxies 
such as their orbital structure, 
total gravitational potential, and phase-space distribution function (DF).
As tautly summarized by \citet[hereafter \citetalias{st96}]{st96}, 
different techniques to create made-to-measure systems 
reproducing the observational data have been devised,
and can be broadly grouped in DF-based, moment-based, orbit-based, and particle-based methods.

DF-based methods fit observations with parametrized functions 
of the integrals of motion or of the action integrals of orbits.
Applications include spherical or integrable systems  
\citep[\eg][]{dejonghe86,dejonghe88,gerhard91,hunterdezeeuw92,carollo95,kronawitter00},
axisymmetric models \citep[\eg][]{hunqian93, dehger94,
kuijken95, mago95, merritt96},
and nearly integrable potentials \citep[\eg][]{dehger93, matger99,binney10}.
The main advantage of these methods is that they provide
the phase-space DF directly, although they generally require
assumptions on the symmetry of the target galaxy.

Moment-based methods find solutions of the Jeans equations
that best reproduce observed quantities 
such as surface density and velocity dispersion
\citep[\eg][]{young80,binmam82,binney90,magobin94,lokas02,cappellari08,williams09,cappellari09}.
Among the drawbacks of these methods are the need for assumptions to close the system of equations,
the lack of any guarantee on the positivity of the underlying DF,
and the difficulties in modelling higher order information
such as the line-of-sight velocity distribution (LOSVD).

Orbit-based methods \citep[][]{sch79,sch93} 
compute a large library of orbits in a fixed potential,
and then adjust the weight of each orbit
until the photometry and kinematics of the target galaxy are well matched
\citep[\eg][] {richstone85, rix97,
vandermarel98, cretton99, 
cappellari02, cappellari06, gebhardt03, 
Valluri04, thomas05, thomas09,vandenbosch10}.
Schwarzschild modelling is widely used,
\eg to infer the masses of black holes at the centers of galaxies,
but applications are mostly restricted to axisymmetric systems.
Moreover, the technique requires the computation 
of a large and representative orbit library 
for every new trial potential. 

Particle-based methods for the most part 
work by slowly correcting individual weights of particles as they are evolved
in the gravitational potential \citepalias{st96}, until the $N$-body system reproduces the observational data.
Kinematic and density observables can be used simultaneously in the weight correction 
by minimizing $\chi^2$-deviations between data and particle model
\citep[hereafter \citetalias{dl07}]{dl07}.
Among the main strengths of this so-called made-to-measure (M2M) technique are
its geometric flexibility, the fact that the potential 
can be evolved self-consistently from the particles,
and that there is no need to store an orbit library.

The M2M method was first applied to the Milky Way's bulge and disk in \citet{bissantz04}.
A version modified to model observational data with errors ($\chi^2$M2M)
was implemented in the parallel code NMAGIC by \citetalias{dl07}.
This has been used to investigate the dynamics 
of the outer halos of the two intermediate-luminosity elliptical galaxies
NGC~4697 and NGC~3379 \citep[hereafter \citetalias{dl08,dl09}]{dl08,dl09},
and of the massive elliptical galaxy NGC~4649 \citep{das11}.
More recent implementations of the M2M method can be found 
in \citet{dehnen09}, who proposed a different technique 
for the weight adaptation, and \citet{longmao10}.
A related particle method but with a different way of adjusting 
to the observational constraints is the iterative technique of \citet{rod09}.

M2M techniques are very promising, but relatively unexplored.
It is therefore a natural question whether these particle methods
can actually recover the phase-space DF of a target galaxy if the data uniquely specify it.
For a given set of data, how much does the final particle model
depend on the initial one?  And how is this dependence influenced by incomplete or noisy data?
Furthermore, given that a system of $N$ particles is trained to match 
a much smaller number of observational constraints,
the problem arises of reducing model degeneracies
and preventing the method from fitting the noise in the data. 

The above issues are related and are connected to the concept of regularization.
In standard $\chi^2$M2M practice,
a weight entropy is used to regularize the particle model:
through the entropy function all particle weights are biased 
towards a smooth distribution of predefined priors, 
which are specified together with the initial model,
and thus implicitly contain assumptions about 
the dynamical structure of the target galaxy.
Therefore, unless the dynamical structure of the galaxy is known beforehand,
smoothing with weight entropy makes it difficult
to construct models with strong phase-space gradients,
\eg between near-radial and near-circular orbits.
This is discussed further in Section~\ref{sec:theory}.
A similar effect arises in Schwarzschild models
regularized using maximum-entropy constraints \citep{richtre88},
which tend to isotropize the final DF \citep[\eg][]{thomas05}.

In this paper, we describe a new Moving Prior entropy Regularization (MPR) method
based on the idea of a distribution of particle priors, 
which are computed according to phase-space occupation 
and which evolve together with the adaptation of the particle weights.
The new method minimizes the dependence of the solution 
on the adopted initial particle model, 
and facilitates recovering both a smoother and more accurate DF,
reducing local fluctuations without erasing global phase-space gradients.

The paper is organized as follows. 
In Section~\ref{sec:theory} the basics of 
the $\chi^2$M2M method are laid out,
the main concerns related to the traditional regularization are explained, 
and our implementation of MPR is developed.
In Section~\ref{sec:target} a series of spherical target models
is constructed for testing the M2M method with MPR.
Then, in Section~\ref{sec:test} and~\ref{ssec:real}
we investigate the different roles played by regularization, initial particle model, 
and data quality for recovering the correct galaxy model,
and we show that the true solution can indeed be recovered from sufficiently good data.
Finally, two astrophysical applications are presented in Section~\ref{sec:gal},
where we reconstruct regularized NMAGIC models 
for the elliptical galaxies NGC~4697 and NGC~3379
in their dark matter halos.
The paper closes in Section~\ref{sec:discuss}.


\section{Regularization of particle models}
\label{sec:theory}
In this Section we outline the $\chi^2$M2M method,
and discuss some issues related to its standard (weight entropy) regularization.
An alternative method to regularize M2M particle models
is then presented.
For a more detailed description of the M2M technique
we refer the reader to \citetalias{st96}, \citetalias{dl07}, \citet{dehnen09}, and \citet{longmao10}.

\subsection{Brief review of $\chi^2$M2M technique to model observational data}
\label{ssec:NMAGIC}
The goal of the $\chi^2$M2M method is to evolve an $N$-body system of particles 
orbiting in a potential to make it reproduce the observables of a target galaxy.
The potential can be either fixed and known \textit{a priori},
or time-varying and self-consistently computed from the particles.

Each particle is characterized by its phase-space coordinates 
${\mathbf z}_i= (\mathbf r_i, \mathbf v_i)$
and by a weight $w_i$.
The particles should be interpreted in a probabilistic sense:
they do not represent single stars
but rather phase-space fluid elements \citep[\eg][]{hernquist92}.
If $M$ is the total stellar mass of the system,
then individual particles have masses $m_i=w_i M/\sum_{i=1}^N w_i$.

An observable of a target galaxy characterized by a distribution function
$f(\mathbf z)$ is defined as
\begin{equation}
Y_j = \int K_j({\mathbf z})f({\mathbf z})~\ud^6z,
\end{equation}
where $K_j$ is an appropriate kernel and $\mathbf z = (\mathbf r, \mathbf v
)$ are the phase-space coordinates.

Given a set of observables $Y_j,\ j=1,...,J$, 
including \eg photometry and kinematics,
the particle weights $w_i$ of the $N$-body system 
are evolved until the model observables 
\begin{equation}
y_j(t)=\sum_{i=1}^N w_i K_j \left[ {\mathbf z}_i(t) \right]
\label{eqn:obs}
\end{equation}
agree with the target observables $Y_j$.
Here, the kernel includes a selection function
which ensures that only particles with a direct effect on the observable $y_j$ 
contribute to it.

Commonly, the model observables are replaced by
their time-averaged values
\begin{equation}
\widetilde{y}_j(t)= \alpha\int_0^{\infty} y_j\left(t-\tau\right)\,e^{-\alpha\tau}\,\ud\tau
\label{eqn:myts}
\end{equation}
to increase the effective number of particles contributing to them,
and to reduce temporal fluctuations.

The task of adapting individual weights of orbiting particles 
until the target and the model observables match
is accomplished by solving the set of differential equations 
referred to as ``force-of-change'':
\begin{equation}
\frac{\ud w_i(t)}{\ud t} = \varepsilon w_i(t)\left(\mu \frac{\partial
S}{\partial w_i} - \sum_j \frac{K_j \left[{\mathbf
z}_i(t)\right]}{\sigma(Y_j)} \uDelta_j(t) \right),
\label{eqn:modFOC}
\end{equation} 
where $\varepsilon$ is a small positive constant, 
and the meaning of the other variables is clarified below.

Equation~(\ref{eqn:modFOC}) maximizes the merit function
\begin{equation}
F = -\frac{1}{2}\chi^2+\mu S
\label{eqn:SF}
\end{equation}
with respect to the particle weights $w_i$.
Here
\begin{equation}
\chi^2 = \sum_{j=1}^J \Delta_j^2 
\label{eqn:chi}
\end{equation}
is a statistical measurement of the goodness of the fit
in terms of deviations
\begin{equation}
\Delta_j(t) = \frac{\widetilde{y}_j-Y_j}{\sigma(Y_j)}
\label{eqn:delta}
\end{equation}
between target and model observables,
taking the error $\sigma(Y_j)$ of the target observable into account.

For the regularization functional, the weight entropy
\begin{equation}
S = -\sum_{i=1}^N w_i \log (w_i/\hat{w}_i)
\label{eqn:entropy}
\end{equation}
is a common choice. $S$ is a measure of the plausibility of the model
in terms of the smoothness of the weight distribution 
and thus, indirectly, of the resulting DF, 
and it serves the purpose of regularization
by pushing the particle weights towards some smooth
predetermined weights $\hat{w}_i$, called priors. 
In typical applications the number of particles is much higher 
than the number of data constraints on the particle model;
this intrinsic ill-conditioning of the problem
translates into a large freedom in the weight adaptation,
and results in models fitting the noise in the data.
That is why a simple minimization of $\chi^2$ is not a well-defined procedure 
to determine the model uniquely,
and a certain degree of regularization is necessary.

The balance between regularization
and fit to observational constraints 
in equation~(\ref{eqn:SF}) 
is controlled by the constant $\mu$,
so that generally models with smaller $\mu$ aim for better fits to the data,
but models with larger $\mu$ have smoother DFs.
In practice, the best choice of $\mu$ is case dependent
\citep[see \eg] [\citetalias{dl08}, \citetalias{dl09}]{gerhard98, thomas05},
hinging on the specific properties
of the observational data to be modelled (error bars, scatter, spatial coverage),
the phase-space structure of the galaxy,
and possibly also the adopted initial particle model.

Finally, note that a likelihood term can be added to equation~(\ref{eqn:SF})
to account for the constraints from a sample of discrete velocities,
as derived in \citetalias{dl08}.

\subsection{Issues with standard weight entropy regularization}
\label{ssec:issues}
In the framework of the $\chi^2$M2M method summarized above,
individual particle weights are slowly adjusted
according to a compromise between $\chi^2$,
which pushes them to match the target observables,
and entropy $S$, which instead 
penalises against deviations of the weights 
from the preassigned set $\{\hat{w}_i\}$ of priors;
more precisely, from $\{\hat{w}_i/\mathrm{e}\}$ 
(see equations~[\ref{eqn:SF}] and~[\ref{eqn:entropy}]).

Even though no rule on the choice of the priors exists,
they are traditionally set to $\hat{w}_i=w_0=1/N$
(the ``uninformative'' or ``flat'' priors in Bayesian statistics), 
and the same is done for the individual weights of the initial particle model.
Through the weight adaptation~(\ref{eqn:modFOC}), then,
the standard Global Weight entropy Regularization (hereafter GWR)
encourages a dynamical structure in the particle model
which is similar to that of the initial particle system.
Of course, this bias is stronger 
for larger values of $\mu$,
and wherever the constraining power of the data is smaller, 
\eg in the outer galactic regions.

In practice, smoothing the weights globally 
towards a set of preassigned, flat priors 
through the entropy~(\ref{eqn:entropy})
makes it difficult to reproduce strong phase-space gradients
of the target galaxy,
\eg strongly anisotropic velocity distributions,
unless either the right orbital structure 
is already in place in the initial particle model,
i.e. its dynamics is known beforehand,
or a very small value of $\mu$ is adopted
at the expense of smoothness of the underlying DF.
This was noticed both in \citetalias{dl08} 
(see their Fig.~10) and \citetalias{dl09},
where under-smoothed models
proved necessary to recover 
strong radial anisotropy in their elliptical galaxy models.

However, under-smoothed particle models 
do not represent a proper solution. 
Indeed, sufficient regularization is needed 
not only to prevent the model from fitting
the noise in the data,
but also to oppose fluctuations of the particle weights caused by the noise in the data,
and to ensure that the weight distribution on neighbouring phase-space tori remains continuous,
as intuitively expected for a relaxed stellar system.

In what follows we present a new regularization method 
which alleviates the main issues of the standard global weight entropy smoothing, 
and permits smooth M2M particle models to be obtained
that reproduce the phase-space gradients of the target galaxies
independently of initial conditions.

\subsection{Alternative regularization based on moving priors}
\label{ssec:idea}
The logical step forward to ease the issues
with the GWR is to abandon the idea of constant priors defined along with 
the initial particle distribution.
Instead we will determine a smooth distribution of particle priors 
which follows the phase-space structures 
traced by the weight distribution, as the weight distribution
evolves to match the observational data.
Then we will use the weight entropy 
to bias particle weights towards such moving priors.

This procedure, which we will denote as Moving Prior entropy Regularization (MPR)
should facilitate a smooth DF without erasing larger-scale phase-space gradients.
In terms of orbits, this means that the new regularization
should assign the same prior to particles 
belonging to the same orbit, 
similar priors to particles on nearby orbits,
and different priors to particles moving 
on very different orbits,
as required in the presence of strong velocity anisotropy.

\subsubsection{Assignment of new local priors}
\label{sssec:loc_priors}
Therefore, based on Jeans' theorem \citep[\eg][]{bt08},
the assignment of new individual priors 
which mirror the underlying evolving DF
is best based on the integrals of motion, respectively orbits, of the particles.
This is particularly simple in the spherical case, 
where the integrals of motion are known
and can be easily found from the particle model.
As already pointed out by \citetalias{dl09}, 
the need for regularization is also
strongest for spherical models,
due to their larger number of independent orbits
with respect to less symmetric systems.
The aim of the present paper is to show that this method works,
and how well it works, in the spherical case.
A simple axisymmetric scheme is shown in Section~\ref{sec:gal},
and generalizations to more complicated geometries 
are sketched out in Section~\ref{sec:discuss}.

For assigning priors in the spherical case,
in practice we sort the particles according to their
energy $E$ and total angular momentum into a rectangular $(E,x)$ grid,
where the so called circularity integral $x=L/L_c$ is the ratio 
between the actual angular momentum
and the angular momentum $L_c$ which
a circular orbit would have at the given energy $E$.
Once particles are binned in a grid of $n_E\times n_x$ 
energy and circularity cells,
we compute the average weight $\hat{w}_{kl}(k=1,\ldots,n_E,l=1,\ldots,n_x)$ 
contained in each cell,
and then we assign it as a new prior 
to all the particles belonging to that cell.

Provided the $(E,x)$ grid correctly resolves 
the relevant phase-space properties of the target,
such new priors ensure an orbit-based regularization
which acts locally, homogenizing the weights of particles moving 
on the same and on neighbouring orbits,
but at the same time tolerates global differences 
among particles on different orbits.

\subsubsection{Smoothing of the grid of particle priors}
\label{sssec:smooth_priors}
We will see that the priors computed in this way can be quite noisy.
To avoid coarseness in the distribution of priors,
and so ensure the global smoothness of the underlying model DF
represented by the $(E,x)$ grid,
we implement a two-dimensional spline fit of the gridded priors.
The technique \citep[][]{press92} is well known and widely used,
also in an astrophysical context
\citep[\eg][]{merritt93, gerhard98, das10}:
a thin-plate spline function for the new priors on the grid is searched,
that minimizes the penalized least square function
\begin{equation}
 \Delta^2\equiv\sum_{k,l}\xi^2 + \lambda\sum_{k,l}\Lambda(\hat{W})_{kl},
 \label{eq:smooth_grid}
\end{equation}
having defined a function $\hat{W}(E,x)$
which equals the values of the priors on the $(E,x)$ grid.
In the equation above, $\xi^2$ measures the deviation 
between the original value of the prior and its spline value in each $(k,l)$-cell, and 
\begin{equation}
\Lambda(\hat{W})_{kl}=\left[\left(\frac{\partial^2\hat{W}}{\partial E^2}\right)^2+2\left(\frac{\partial^2\hat{W}}
{\partial E\partial x}\right)^2+\left(\frac{\partial^2\hat{W}}{\partial x^2}\right)^2\right]_{\substack{E=E_k\\x=x_l}}
\end{equation}
quantifies the complexity of the fitting spline
in terms of the second derivatives, 
which are numerically computed via finite differences.

The regularization parameter $\lambda$
determines which of a family of splines,
ranging from a plane for $\lambda\to\infty$ 
to an interpolating cubic spline surface for $\lambda\to0$,
is fitted to the grid of priors.
Obviously, the optimal $\lambda$ is that which resolves
the relevant structures in the underlying prior distribution,
but at the same time damps strong and presumably spurious variations among nearby priors.

In principle, $\lambda$ can be calibrated
with a sequence of experiments on the $(E,x)$ grid.
However, since particle weights evolve in time,
and so does the grid of priors,
we decided to implement the General Cross Validation technique \citep[GCV, ][]{wahba90}
to determine automatically the optimal value of $\lambda$ 
each time a new grid of priors is computed.
GCV is based on the principle 
of sequentially omitting each data point, re-fitting the spline,
and predicting the value of the point from the spline.
The technique singles out the optimal value of $\lambda$ for this to work best. 

\subsubsection{New definition of pseudo-entropy}
\label{sssec:new_S}
The new moving priors substitute the traditional ones
in the definition~(\ref{eqn:entropy}) of the pseudo-entropy,
which we slightly modify in order to account for the normalization of the weights.
As already noted, maximizing the standard entropy
biases the weights towards $\hat{w}_i/\mathrm{e}$,
so that oversmoothing (\eg for high values of $\mu$)
causes an undesired global decrease of all weights
which leads to a poor fit of the mass distribution 
(see \eg Fig.~10 in \citetalias{dl08}).

In order to avoid such problems, we define a new weight entropy
\begin{equation}\label{S+}
S=-\sum_{i=1}^N w_i\left[\log\left(\frac{w_i}{\hat{w}_i}\right)-1\right],
\end{equation}
for which we can immediately check that $(i)$ maximizing
this quantity pushes the weights to the actual values of the priors,
$(ii)$ positive and negative corrections to the weights are now a priori equally likely,
and $(iii)$ the whole regularization scheme is neutral to mass,
so that the only power to alter the total mass of the system is left to the data, 
see Section~\ref{ssec:obs1}.

\section{Target models and observables}
\label{sec:target}
In this Section we construct a series of spherical targets
to be modelled with NMAGIC (Section~\ref{sec:test} and~\ref{ssec:real})
in order to address two issues,
namely $(i)$ testing the ability of the new regularization scheme
to fit the target data with an intrinsically smooth 
and unbiased particle model,
and $(ii)$ exploring the extent to which the $\chi^2$M2M technique 
can recover the target phase-space structure from a given data set
independently of the initial particle model.

With these aims in mind, we first focus on 
a problem whose solution is theoretically known to be unique.
As proved by \citet{dejonghe92} in the spherical non-rotating case, 
if the gravitational potential is known
and complete information on the LOSVD at all radii is available,
then the underlying DF can be uniquely recovered.
Therefore, the first target model we design has a known spherical potential 
and is truncated in radius, so that photometric and kinematic data 
can fully constrain it.

As a second target model, we build an untruncated (infinite) system
whose outer regions remain unconstrained by data,
similar to the case of modelling real galaxies.

For both target models, we use the 3D luminosity density
together with the LOSVD along several long-slits as target data in the modelling,
similar to \citetalias{dl08} and \citetalias{dl09}.
For each model, we generate both
a set of idealized kinematic data, i.e. a large number of data points
with small error bars, and a set of more realistic, i.e. sparser and noisier, data points.
We use NMAGIC itself to construct our target dynamical equilibrium structures, 
and to determine their observables,
as described in more detail in the following subsections.

\subsection{Spherical anisotropic Hernquist targets}
\label{ssec:target}
Our target models are \citet{hernquist90} spheres
with a radially anisotropic orbital structure 
either of the Osipkov-Merritt kind \citep[][hereafter OM]{osipkov79, merritt85},
or of the more mildly anisotropic, quasi-separable kind \citep{gerhard91}.
Generally speaking, they are isotropic in the central regions,
and radially anisotropic for radii
greater than a specified anisotropy radius.

The potential-density pair for Hernquist models is
\begin{equation}
\rho (r) = \frac{a M}{2 \pi r(r+a)^3}, \qquad
\varphi (r) = -\frac{G M}{r+a},
\label{eqn:her}
\end{equation} 
where $M$ is the total mass, $G$ the gravitational constant
and $a$ the scale length.
We set the scale length equal to $1$ kpc,
and we use characteristic values of the elliptical galaxy NGC~3379
for the total luminosity $L=1.24\times10^{10} L_{\sun}$,
the stellar mass-to-light ratio $\Upsilon=5$,
and the distance $D=9.8$ Mpc.
The projected effective radius of our target model is
$R_{\rm eff}\approx 38.3''=1.82$ kpc.

With respect to the orbital anisotropy,
we either fix the OM anisotropy radius $r_a=2a$,
or we use $\alpha=2$ and $L_0=0.3\sqrt{GMa}$ 
in the prescription of \citet{gerhard91} to generate 
moderately radially anisotropic models
(see equations~[2.2] and~[3.14] therein).

Following the method described in \citet{debatsel00},
we generate particle model realizations of the spherical targets.
To construct a truncated target, we only retain particles
with energies lower than $E_{\rm max}\equiv\varphi(r_{\rm max})$, with $r_{\rm max}$
equal to the model boundary.

Finally, we relax the particle models in the fixed Hernquist potential 
(note that the truncated target is therefore not a self-consistent system), 
and we compute the target observables from the final particle model using NMAGIC 
to integrate the particles, as detailed below.

\subsection{Luminosity observables}
\label{ssec:obs1}

We consider as density constraint
a spherical harmonics expansion of the target luminosity density 
on a 1-D mesh of radii $r_k$. The expansion coefficients
\begin{equation}
a_{lm,k} = L \sum_i \gamma_{ki}^{CIC} Y_l^m(\theta_i,\varphi_i) w_i
\label{eqn:alm}
\end{equation}
are computed from the particle realizations through NMAGIC, 
making use of the cloud-in-cell technique \citep[see \eg \citetalias{dl07},][]{bt08}
to distribute the weight of a particle between nearby grid points.
In the definition above, $L$ is the total luminosity of the target, 
$Y_l^m$ are the standard spherical harmonic functions,
and $\gamma_{ki}^{CIC}$ is the selection function
associated with the cloud-in-cell scheme.
The radial grid has 60 points,
quasi-logarithmically spaced between $r_{\rm min}=0.01''$ 
and $r_{\rm max}$ equal to the model boundary (for the truncated target)
or to $1500''\sim40 R_{\rm e}$  (for the infinite target).

Poissonian error bars, dependent on the number of particles in each shell, 
are assumed for the radial mass, 
while 50 Monte-Carlo realizations 
of the density field of the target model
allow errors to be assigned to the higher order mass moments (see \citetalias{dl07}).
Because the targets are spherical, all model $a_{lm,k}$
with $l\neq0, m\neq0$ are constrained to be zero within these errors,
while the $a_{00,k}$ are constrained by their values
for the known target luminosity distribution.

When comparing the target data with the model observables,
we compute the latter in the exact same way from the particle model.

By fitting the $a_{lm,k}$ coefficients~(\ref{eqn:alm}), the total luminosity
of the model is adjusted to the target luminosity L. The sum of the weights,
initially set to $\sum_{i=1}^Nw_i=1$, may therefore change if the luminosity of the initial model
$L_{\rm initial}\neq L$. In this work, we set $L_{\rm initial}=L$,
and we do not adjust the mass-to-light ratio, except in Section~\ref{sec:gal},
so that the total mass is also constant throughout the evolution.

\subsection{Kinematic observables}
\label{ssec:obs2}
As kinematic target observables, we use the luminosity-weighted Gauss-Hermite moments 
of the LOSVD \citep{vandermarel93,gerhard93}
in various slit cells, computed from the particle realizations using NMAGIC, through
\begin{equation}
b_{n,p} \equiv l_p\, h_{n,p} = 2\sqrt{\pi}L\sum_i \delta_{pi}
u_n(\nu_{pi}) w_i
\label{eqn:lumghn}
\end{equation}
(\citetalias{dl07}).
Here, $l_p$ is the luminosity in slit cell $\mathcal{C}_p$,
$\delta_{pi}$ selects only particles belonging to that cell, 
the dimensionless Gauss-Hermite functions are
\begin{equation}
u_n(\nu) = \left(2^{n+1}\pi n!\right)^{-1/2}H_n(\nu)\exp \left(-\nu^2/2\right),
\end{equation} 
where $H_n$ denote the standard Hermite polynomials,
and finally
\begin{equation}
\nu_{pi}=\left(v_{z,i}-V_p\right)/\sigma_p,
\end{equation}
with $v_{z,i}$ the line-of-sight velocity of particle $i$,
and $V_p$ and $\sigma_p$ the best-fitting Gaussian parameters 
of the target LOSVD in the given slit cell.
\begin{figure}
\includegraphics[angle=-90.0,width=1.\hsize]{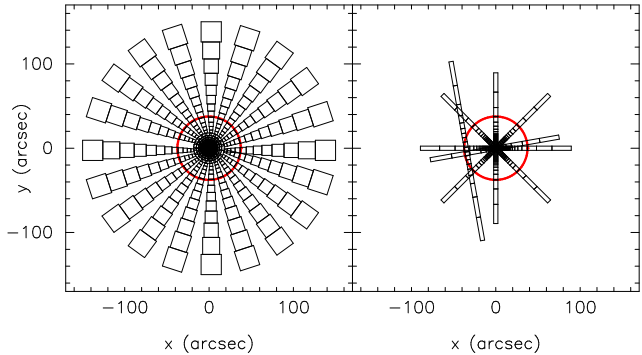}
\caption[]{Geometry of the two different slit setups.
The red circle corresponds to $R_{\rm{eff}}$.
\textit{Left panel:} \textsl{idealized} configuration of 10 slits
covering the target; slit cells outside $R_{\rm{eff}}/2$ are square (see Section~\ref{ssec:obs2}). 
\textit{Right panel:} \textsl{realistic} slit configuration, 
adapted from NGC~3379 (\citetalias{dl09}).}
\label{fig:slit_geo}
\end{figure}

To compute $V_p,\sigma_p, h_3,h_4$
from the particle model, we adopt the following procedure.
First, we compute the mean velocity and rms velocity 
of particles in each slit cell, and use them to estimate
the $b_{n,p}$ from the particles through equations~(\ref{eqn:lumghn}).
Next, we use the first order relations
\begin{equation}
\Delta h_1 = -\frac{1}{\sqrt{2}} \frac{\Delta V}{\sigma}; \quad
\Delta h_2 = -\frac{1}{\sqrt{2}} \frac{\Delta \sigma}{\sigma}
\label{eqn:h1h2}
\end{equation} 
\citep{vandermarel93, rix97} iteratively
to correct $V_p$ and $\sigma_p$, until $h_1$ and $h_2$ both converge to zero. 
Then, the new target moments $b_{n,p}$ are temporally smoothed
to reduce fluctuations caused by particle noise,
and this leads to values of  $h_1$ and $h_2$
slightly different from zero.  
Finally, the resulting velocity profile is fitted by a Gauss-Hermite series \citep{vandermarel93} 
setting $h_1=h_2=0$, and the $b_{n,p}$ are recomputed.

Two different slit configurations are considered, 
as shown in Fig.~\ref{fig:slit_geo}.
There, the left panel illustrates a schematic view
of an idealized slit data setup, which consists of 10 slits
covering the target and extending as far as $r_{\rm max}=150''\sim4 R_{\rm eff}$.
In order to increase spatial coverage in the outer regions,
and so to decrease the effects of particle noise,
slit cells outside $R_{\rm{eff}}/2$ are enlarged and made square.
Moreover, Gauss-Hermite coefficients up to $h_6$ are considered.
To assign error bars to the target kinematic data,
we compute averaged values of the final time-smoothed $b_{n,p}$ moments 
in the 10 different slits, and then we set 
the errors equal to $\sqrt2$ times 
the rms deviation of the individual slit cell moments from the average\footnote{The factor 
$\sqrt2$ in the error bars is necessary because,
given this generation of kinematic data, 
the NMAGIC model will have an intrinsic particle noise  
similar to that of the data which it will try to match.}.
To complete the generation of this slit data set, 
Gaussian random variates with $1\sigma$ equal
to these errors are added to the average moments $b_{n,p}$. 
In the following, we refer to this kinematic data set 
as \textsl{idealized} data.

For the truncated model, these data are sufficiently close
to the required ``complete'' data set that we would expect
to be able to recover the theoretically unique underlying model
with very good accuracy.

The right panel of Fig.~\ref{fig:slit_geo} shows instead the 6 slits
which were used by \citetalias{dl09} to model NGC~3379;
for this second slit configuration, $r_{\rm max}=100''\sim3 R_{\rm eff}$,
only $v,\sigma, h_3,h_4$  are available,
and observational errors for this galaxy are adopted.
Finally, Gaussian random variates are added to the data
with $1\sigma$ equal to the observational errors.
Hereafter we refer to this latter kinematic data set 
as \textsl{realistic} data.

\section{Convergence to a theoretically unique solution}
\label{sec:test}

Here we construct NMAGIC models
for the radially anisotropic target galaxy model described above.
As constraints we use the luminosity density
and the \textsl{idealized} kinematic data.
We determine the amount of regularization, 
investigate whether it is possible to converge 
to the theoretically unique solution, 
and see how well the target galaxy can be reproduced
starting from different initial particle models.

Our NMAGIC models show for this case that if a unique inversion of data 
to recover the underlying target DF exists,
then it can actually be found via $\chi^2$M2M modelling from good enough data.
The new regularization method proposed in this paper
significantly improves both the accuracy 
with which the target intrinsic properties are reproduced,
and the convergence to the right solution
independently of initial conditions.

\subsection{Modelling procedure and diagnostic quantities}
\label{ssec:runs}
Starting from an initial particle model, 
the weights of all particles are evolved 
until the particle system matches the target. 
As initial particle system, we adopt an isotropic
Hernquist sphere with the same luminosity but scale length $a=1.5$ kpc.
Different velocity distributions are also considered, as specified below.
During the whole evolution, the potential is kept fixed
to the target potential.
The particles are integrated using a leap-frog scheme.

After a relaxation phase in which the particle system is advanced
without weight correction, 
weights are updated according to the force-of-change in equation~(\ref{eqn:modFOC}), 
i.e. subject to both data constraints and smoothing constraints,
for $\sim10^5$ correction time steps.
We define the model to have converged 
if $\chi^2/J$ averaged over 50 steps is almost constant in the last $10^4$ steps, 
with fluctuations which are typically of order 2\%.
The particle weights are then constant to a similar accuracy with MPR.
Finally, the particles are freely evolved
for another $10^4$ steps without any further weight correction,
to ensure that the final particle model is well phase-mixed.
For reference, $10^4$ correction time steps correspond to $\sim42$
circular rotation periods at the target $R_{\rm eff}$.

When using the new regularization scheme,
individual priors are not kept constant in time
but rather they are continuously updated
while particle weights are changed to match the target observables,
as detailed in Section~\ref{ssec:idea}.
Particles are sorted according
to their orbital integrals in a grid of $n_E=30$ and $n_x=10$ bins,
chosen as a compromise between retaining good resolution for the orbit distribution 
and ensuring a sufficient number of particles in all grid cells.
The average weight contained 
in each grid cell is computed, and then
a GCV thin-plate smoothing spline is fitted
to the distribution of average weights on the grid.
The spline value in every grid cell is finally assigned 
as the new prior to all particles belonging to the cell.

\begin{figure}
\includegraphics[angle=-90.0,scale=0.45]{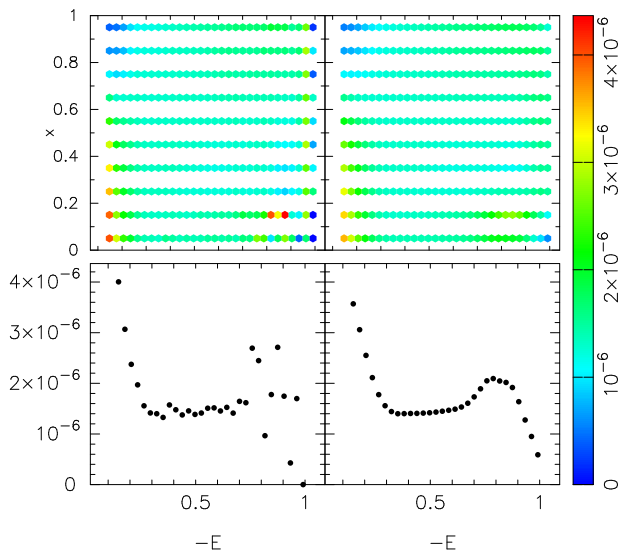}
\caption[]{Top: unsmoothed (\textit{left}) and smoothed (\textit{right}) grid of
individual particle priors after $\sim10^3$ correction time steps (colour bar on the right).
Bottom: a cut of the above grid for $x=0.05$,
showing unsmoothed (\textit{left}) and smoothed (\textit{right}) 
particle priors as a function of energy.
Priors are smoothed among nearby cells with the GCV thin-plate smoothing spline
described in Section~\ref{sssec:smooth_priors}.
}
\label{fig:smooth_grid}
\end{figure}

A typical outcome of the procedure early in the evolution 
is shown in Fig.~\ref{fig:smooth_grid}, where the $(E,x)$ grid of priors
is plotted before (left) and after (right) smoothing,
and similarly for a horizontal cut 
(fixed angular momentum) through the grid.
The cut shows considerable noise before smoothing 
in the central grid cells,
where even with a total $N\sim10^6$ particles
the number of particles per cell for a Hernquist model cusp is still small.
In the tests presented here, new priors are computed 
frequently in the initial phase
and every $10^4$ correction time steps later in the run, 
which results in an efficient regularization at a minimum computational cost.
When testing the new scheme for very large $\mu$ values,
priors are computed and updated more often. 

The quality of the final particle model of each run is assessed
through three diagnostic quantities.
The first is the goodness of the fit to the data
in terms of $\chi^2/J$, 
where $J$ is the number of data points.
Assuming the goodness of fit statistics 
follows a $\chi^2$ probability distribution function,
the mean of the $\chi^2$ distribution is equal 
to the number of degrees of freedom,
i.e. the number of constraints (data points plus constraints introduced by the merit function)
subtracted by the number of parameters (model parameters plus fitted weights),
which are both difficult to quantify.
However, if the number of degrees of freedom is approximately
equal to the number of data points, 
then $\chi^2/J<1$ means that we are fitting the data well.

The second is the level to which the known intrinsic kinematics 
of the target galaxy are recovered by NMAGIC,
quantified by the rms difference between the intrinsic velocity moments of the target galaxy
and those of the final particle model realization.
In the following, the internal kinematics (streaming velocity and velocity dispersions)
of the particle model are computed by binning particles in spherical polar coordinates,
using 21 radial shells quasi-logarithmically spaced 
between $r_{\rm{min}}=0.01''$ and $r_{\rm{max}}$, 
12 bins in the azimuthal angle $\phi$, 
and 21 equally spaced bins in $\cos\theta$,
where $\theta$ is the polar angle.

Finally, we determine the degree to which the particle model
reproduces the known phase-space structure of the spherical target.
To quantify this we compute 
the mass-weighted relative rms difference between model and target weights 
($w_{kl,m}$ and $w_{kl,t}$, respectively)
in the grid of energy and circularity $(E,x)$
used also for the regularization:
\begin{equation}
\Delta_{\rm grid}=\sqrt{\sum\limits_{k,l}w_{kl,t}\left(\frac{w_{kl,t}-w_{kl,m}}{w_{kl,t}}\right)^2/\sum\limits_{k,l} w_{kl,t}}.
\end{equation}

\subsection{Calibrating regularization}
\label{sssec:mu}
Following the same approach as \citet{gerhard98}, \citet{thomas05},
\citetalias{dl08}, and \citetalias{dl09}, we construct NMAGIC models
for the target galaxy which only differ 
in the adopted regularization scheme and the amount of regularization,
i.e. the value of the parameter $\mu$.
Note that $\varepsilon$ is kept constant between all models.
\begin{figure}
\includegraphics[angle=-90.0,width=0.95\hsize]{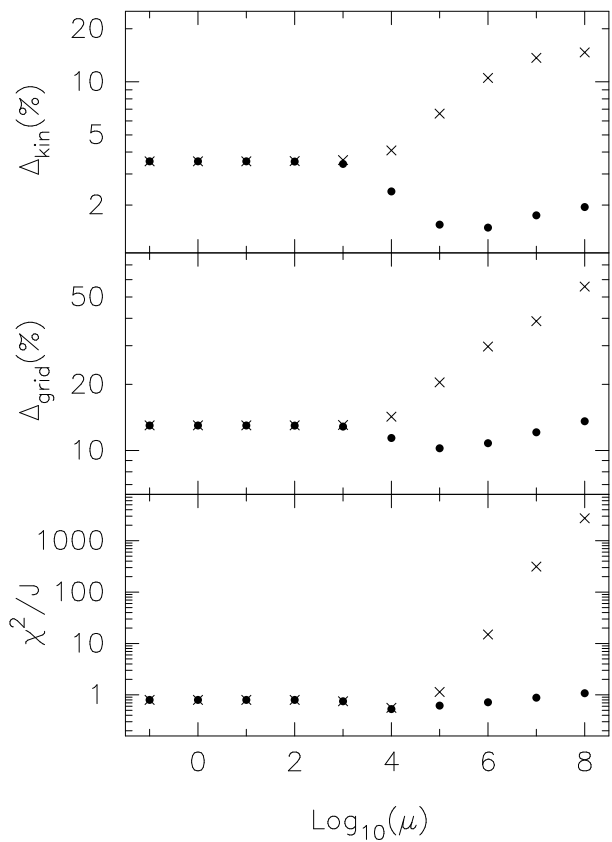}
\caption[]{Quality of the final NMAGIC particle models
as a function of the regularization parameter $\mu$.
\textit{Top panel}: rms deviation (\%) of the final particle model from the target 
internal velocity moments. \textit{Middle panel}: rms deviation (\%)
between the occupation of the $(E,x)$ grid of the target and of the final NMAGIC model.
\textit{Bottom panel}: goodness of the particle model fit
to photometric and kinematic data.
Crosses refer to models obtained with the traditional GWR scheme,
dots to models obtained with the new MPR.
}
\label{fig:mu}
\end{figure}

The results are summarized in Fig.~\ref{fig:mu},
where the normalized goodness of fit $\chi^2/J$,
the mass-weighted rms over the $(E,x)$ grid, $\Delta_{\rm grid}$,
and finally the rms difference between the internal velocity moments of the target 
and final particle model, $\Delta_{\rm kin}$, are plotted as a function of $\mu$, 
from unsmoothed models (small $\mu$) to oversmoothed models (high $\mu$).

We first focus on the NMAGIC particle models obtained
with the traditional GWR technique (crosses, Fig.~\ref{fig:mu}).
For a wide range of values of $\mu\leq10^4$,
NMAGIC is able to fit the data with $\chi^2/J\lsim 1$.
No clear minimum is present in the plotted rms deviations in grid and intrinsic kinematics
as a function of $\mu$: these quantities stay almost flat for a large range of $\mu$,
and then rapidly increase for $\mu\gsim10^4$,
when the increasing amount of smoothing upsets the fit to the data.
By the time the smoothing becomes effective in damping fluctuations
in the intrinsic quantities, the bias introduced by the global nature
of the smoothing has already set in - hence no clear optimal value of $\mu$
is found.
For GWR and this particular data set,
$\mu=10^4$ gives a good compromise between quality of the data fit
and recovery of the target properties - but with little smoothing.

\begin{figure}
\includegraphics[angle=-90.0,scale=0.47]{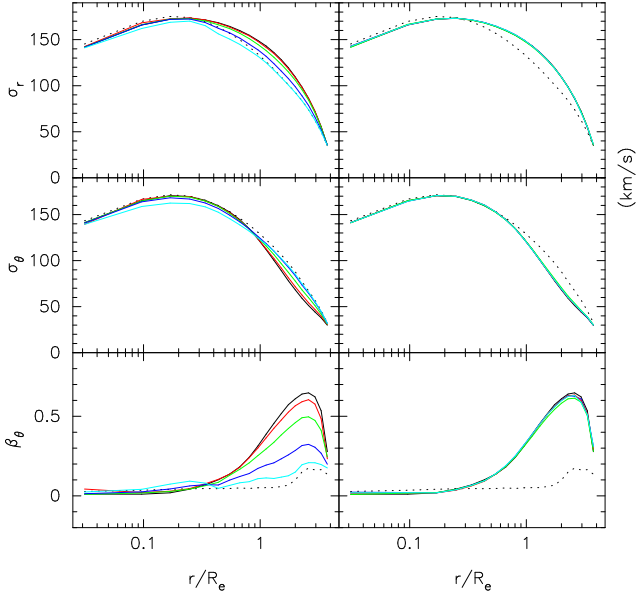}
\caption[]{Intrinsic kinematics of the NMAGIC models
obtained with the GWR (\textit{left panel}), 
and with the new MPR method (\textit{right panel}).
From top to bottom: radial velocity dispersion profile,
vertical velocity dispersion profile, and anisotropy parameter.
The dotted and full black lines show the intrinsic kinematics
of the initial near-isotropic particle model, and of the truncated target galaxy, respectively.
Red, green, blue, and light blue lines correspond to $\mu=10^3,10^4,10^5,10^6$ adopted in the modelling 
(see Section~\ref{sssec:mu}).}
\label{fig:trunc_int}
\end{figure}

How well the intrinsic kinematics of the target galaxy can be recovered
is shown in the left panel of Fig.~\ref{fig:trunc_int}, 
which compares the known target kinematics with the final NMAGIC models
obtained with $\mu=10^4,10^5,10^6,10^7$.
As expected, for higher values of $\mu$ the internal moments
remain closer to the initial isotropic moments.

\begin{figure}
\includegraphics[angle=-90.0,width=0.98\hsize]{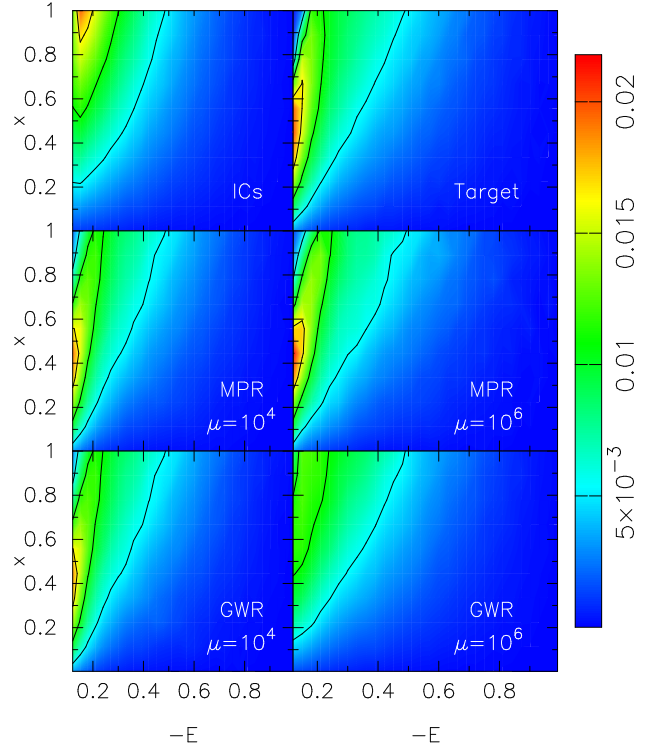}
\caption[]{Mass distribution function (MDF) of particle weights in the $(E,x)$ grid,
for the initial particle model (top left), the truncated target galaxy (top right),
the models obtained with the standard GWR (bottom)
and with the MPR method (middle), for different values of $\mu$.
The colour scheme reflects the total weight contained in each grid cell,
where $n_E=30$, $n_x=10.$}
\label{fig:grid}
\end{figure}

Fig.~\ref{fig:grid} shows the level to which the known target DF can be recovered by NMAGIC
for $\mu=10^4,10^6$:
the distribution of total particle weights
in the $(E,x)$ grid is plotted for the initial particle model,
the truncated target, and the models obtained with NMAGIC.
We denote this by ``mass distribution function'', or MDF for brevity.
Clearly, for $\mu=10^4$ the main phase-space structures are well recovered,
showing that NMAGIC is able to fit the data
and to approximately find the underlying MDF,
but the peak on high-E near-radial orbits is underestimated 
because of the global nature of GWR.
For the more heavily smoothed case with $\mu=10^6$,
this peak is completely wiped out.

We now consider $\chi^2$M2M models obtained by fitting the same target data
with MPR.
As can be seen in Fig.~\ref{fig:mu} (black dots),
the new method works very well in reproducing the target,
and a series of NMAGIC models fitting 
the photometric and kinematics constraints of the galaxy 
within $\chi^2/J\lsim 1$ can be generated
for a wider range of $\mu$ values.
Of these, models obtained for values of $\mu$
$\lsim10^3$ are essentially driven by 
the $\chi^2$ term alone.
However, when regularization becomes significant,
a minimum is reached
both in the rms deviation between intrinsic moments of the particle model and of the target,
and in the rms deviation of their $(E,x)$ distributions
(top and middle panels of Fig.~\ref{fig:mu}, respectively).
Remarkably, the minimum is well below 
that achievable with traditional weight entropy smoothing,
indicating that the phase-space structure
and the internal moments of the target
can be recovered much better using MPR.

The right panel of Fig.~\ref{fig:trunc_int}
shows how close the internal kinematics of the final particle models
for $\mu=10^4,10^5,10^6,10^7$ are to those of the target galaxy.
Note that the residuals are so small that the trend 
with $\mu$ seen in Fig.~\ref{fig:mu} cannot be seen;
the new scheme allows one to recover the target moments almost perfectly.
The accuracy with which the MDF in $(E,x)$ integral space is reproduced
is shown in Fig.~\ref{fig:grid}.
Visually comparing this plot with the corresponding ones for the truncated target
and the best model obtained using standard GWR, 
shows that the target is now recovered much better.
In particular at small energies, i.e. in the outer regions,
the weight of particles on radial orbits is increased
while that of particles on circular orbits is decreased
more effectively with MPR, especially for the preferred $\mu=10^6$.

We conclude that, for this particular data setup, 
the best choice for $\mu$ with MPR is $\sim10^6$.
This value is considerably larger than the corresponding $\mu$ of the traditional GWR,
showing that the new regularization
succeeds better in reconciling the smoothness of the underlying model with orbital anisotropy. 

\begin{figure}
\includegraphics[angle=-90.0,scale=0.4]{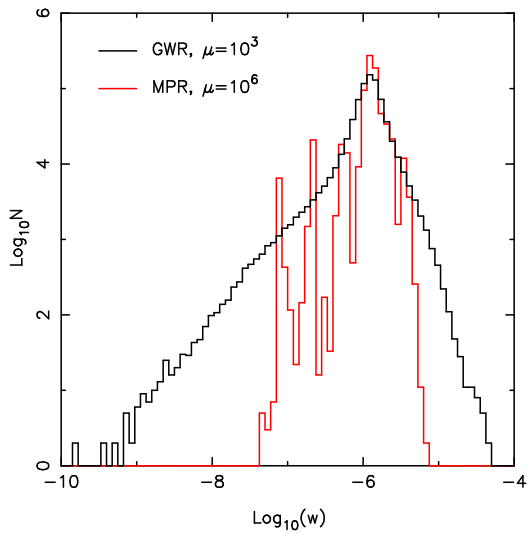}
\caption[]{Distribution of particle weights for the final
optimally smoothed NMAGIC models obtained with the traditional GWR
(black histograms, $\mu=10^4$) and with the MPR scheme (red histograms, $\mu=10^6$).
Particle weights were initialized to $w_0=1/750000\sim10^{-6}$.
}
\label{fig:wdist}
\end{figure}

It is instructive to compare the final distribution
of particle weights for both regularization schemes.
Fig.~\ref{fig:wdist} shows that MPR results 
in a more compact and more structured weight distribution, which avoids extended tails
of extremely increased or decreased weights,
while still providing a good and less noisy fit the data (see below).
A similar comparison in the context of Schwarzschild modelling
can be found in Fig.~17 of \citet{thomas07}.

\begin{figure}
\includegraphics[angle=-90.0,scale=0.4]{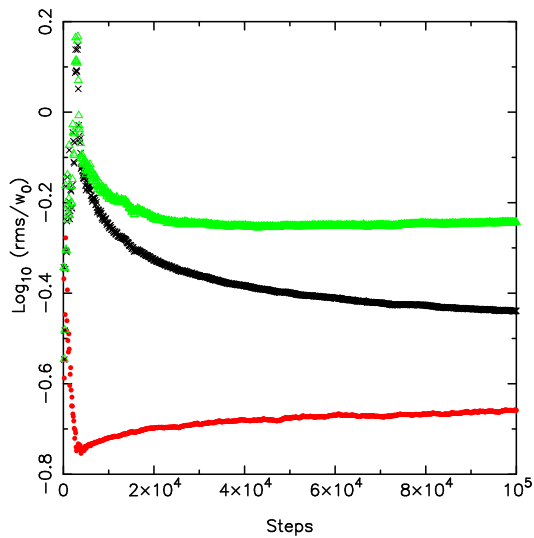}
\caption[]{Rms fluctuations of particle weights around the mean value in each $(E,x)$ cell
for the optimally smoothed NMAGIC models obtained with GWR
(black crosses, $\mu=10^4$) and with MPR (red dots, $\mu=10^6$),
as a function of the number of NMAGIC correction time-steps.
The top curve (green triangles) shows an essentially unsmoothed model ($\mu=10^3$).
$w_0=1/750000\sim10^{-6}$ is the value of the initial particle weights.
The grid has $n_E=30$ times $n_x=10$ cells, and only those containing more than 50 particles
are taken into account, to avoid particle noise effects in the computation of the residuals.
}
\label{fig:res}
\end{figure}

Along with a more compact weight distribution, MPR also leads
to a smoother particle model.
This can be quantified by computing the rms fluctuations of particle weights 
around the mean value in all the cells
of the $(E,x)$ grid for the different kinds
of regularization, as shown in Fig.~\ref{fig:res}.

Not unexpectedly, the fit to the data also looks smoother 
when adopting the new regularization,
and the larger $\mu$ value permitted by this scheme 
opposes an overfitting of the data points,
as can be appreciated for different observables in Section~\ref{sec:gal}.

To summarize, we have tested the $\chi^2$M2M method
with a new Moving Prior Regularization scheme for a radially anisotropic
truncated target model with \textsl{idealized} data,
and have calibrated the best value of the regularization parameter $\mu$.
We have shown that the corresponding NMAGIC models match the target data well
and recover the MDF for this model in its known potential.
We have also seen that these models are much less sensitive
to the value of $\mu$ than with the traditional weight entropy regularization,
which can only reproduce the global anisotropy of this model essentially without smoothing.
The new regularization scheme allows NMAGIC to recover a
particle model that fits the data well but is both intrinsically smoother
and reproduces the properties of the target more accurately.

\subsection{Varying the initial particle model}
\label{sssec:ics}
The results of the previous subsection already show 
that NMAGIC can recover the orbit distribution of our truncated spherical target galaxy 
from a set of data that specifies it essentially uniquely.  
In this experiment, we used an isotropic initial model, 
so now we investigate the natural question whether and how 
this result is dependent on the choice of initial particle model. 

In particular, we consider both the case
in which the initial particle model 
has a radially anisotropic OM orbital structure
(with anisotropy radius $r_a=3a$, different from the target galaxy),
and the case in which it is tangentially anisotropic according 
to the quasi-separable prescription of \citet[][with $\alpha=2$, $L_0=0.3\sqrt{GMa}$, $c=0.1$]{gerhard91}.
For both initial particle models we checked that 
a minimum number of particles on radial and tangential orbits is present at each energy.
By analogy with the experiment described in the previous subsection,
the initial particle models are Hernquist spheres
with scale-length $a=1.5$ kpc, larger than the target galaxy ($a=1$ kpc).
The same setup of the NMAGIC run is adopted,
together with the optimal $\mu$ values determined in Section~\ref{sssec:mu} above
for the two regularization methods.

\begin{figure}
\includegraphics[angle=-90.0,width=0.98\hsize]{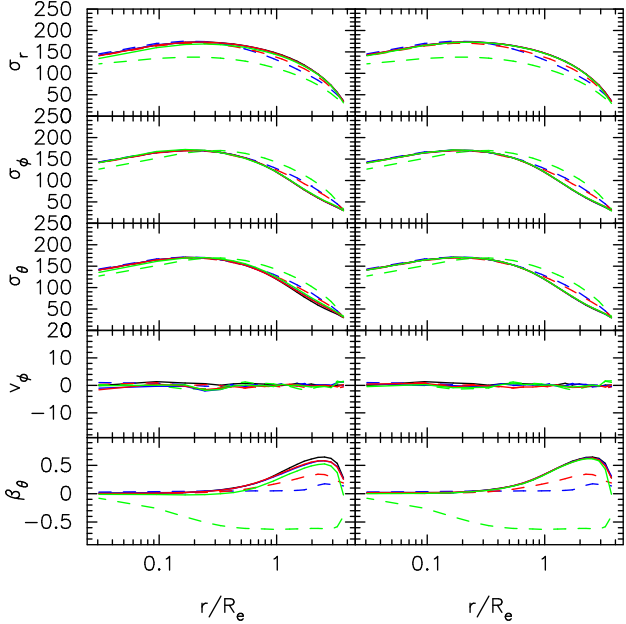}
\caption[]{Truncated target with \textsl{idealized} data:
recovery of the intrinsic kinematics with different initial particle models.
From top to bottom: radial, azimuthal, 
and vertical velocity dispersion profiles,
mean azimuthal streaming velocity, and anisotropy parameter
of the NMAGIC models (full lines) 
for different initial conditions (dashed lines).
The black line indicates the intrinsic moments of the target galaxy.
Blue, red and green colours correspond
to isotropic, radially anisotropic and tangentially anisotropic
initial conditions, respectively.
GWR was adopted in the runs shown in the \textit{left panel},
while MPR in the runs shown in the \textit{right panel}.}
\label{fig:trunc_ics_int}
\end{figure}

\begin{figure}
\includegraphics[angle=-90.0,scale=0.6]{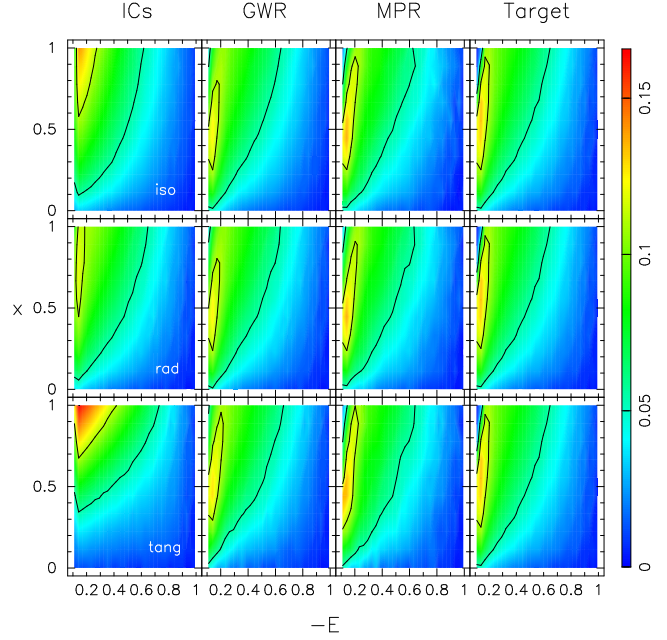}
\caption[]{Recovery of the MDF of the truncated target (last column on the right)
for different initial particle models from \textsl{idealized} kinematic constraints.
The first column shows the distribution of particle weights in the $(E,x)$ grid
for the isotropic, radially and tangentially anisotropic initial particle models (from top to bottom).
The second column corresponds to the final NMAGIC particle models
obtained with traditional GWR;
the third column to the models obtained using the new MPR.
The colour scheme reflects the square root of the total weight contained in each grid cell,
where we have adopted $n_E=30, n_x=10$.}
\label{fig:ics_grid}
\end{figure}

With the new MPR scheme, the final NMAGIC models obtained for different initial orbital distributions
differ remarkably little. 
Table~\ref{tab:trunc_ics} and 
Figs.~\ref{fig:trunc_ics_int}-\ref{fig:ics_grid}
show how well the intrinsic kinematics and phase-space MDF 
match those of the known target galaxy.
The MDF of the final particle model is very similar
to that of the target galaxy, indepentent of 
the choice of initial conditions (Fig.~\ref{fig:ics_grid}).
Typical fluctuations in the mass-weighted relative rms difference 
between target and model MDF are $12\%$,
while the intrinsic kinematics of the target
is recovered almost perfectly, as shown in Fig.~\ref{fig:trunc_ics_int}.

\begin{table}
\centering
\begin{tabular}{l c l l l l}
\hline
& & ICs & $\chi^2/J$ & $\Delta_{\rm kin}(\%)$ &  $\Delta_{\rm grid}(\%)$ \\ 
\hline
\textsl{Idealized} & GWR   & iso  &  0.57  & 4.07 \textit{(14.67)} & 14.24  \textit{(43.98)} \\
data               &       & rad  &  0.46  & 3.62 \textit{(11.85)} & 12.94  \textit{(32.25)}\\
                   &       & tang &  1.64  & 7.56 \textit{(27.20)} & 17.51  \textit{(75.46)} \\
                   & MPR   & iso  &  0.72  & 1.49 \textit{(14.67)} & 10.80  \textit{(43.98)} \\
                   &       & rad  &  0.67  & 1.62 \textit{(11.85)} & 11.21  \textit{(32.25)} \\
                   &       & tang &  1.57  & 3.38 \textit{(27.20)} & 12.53  \textit{(75.46)}\\
\hline
\textsl{Realistic}& GWR   & iso   &  0.26  & 2.44 \textit{(5.17)}  & 7.49 \textit{(17.63)} \\
data              &       & rad   &  0.20  & 3.25 \textit{(3.85)}  & 12.68 \textit{(15.79)}\\
                  &       & tang  &  0.29  & 7.42 \textit{(16.14)} & 20.85 \textit{(56.44)}\\

                  & MPR   & iso   &  0.37  & 1.32 \textit{(5.17)}  & 7.39 \textit{(17.63)} \\
                  &       & rad   &  0.38  & 3.13 \textit{(3.85)}  & 10.27 \textit{(15.79)} \\
                  &       & tang  &  0.38  & 2.35 \textit{(16.14)} & 11.09 \textit{(56.44)} \\
\hline
\end{tabular}
\caption{NMAGIC models for the truncated target galaxy.
Different initial particle models are adopted.
For the traditional weight entropy smoothing $\mu=10^4$;
for the new regularization scheme, $\mu=10^6$. 
The goodness of fit $\chi^2/J$, $\Delta_{\rm kin}$ and $\Delta_{\rm grid}$ are computed as described in Section~\ref{ssec:runs}.
In brackets, the same $\Delta_{\rm kin}$ and $\Delta_{\rm grid}$ computed between
the initial particle model and the target.
}
\label{tab:trunc_ics}
\end{table}

It is instructive to see how a similar result cannot be achieved 
with the traditional weight entropy:
an inspection of Table~\ref{tab:trunc_ics}, or Fig.~\ref{fig:trunc_ics_int} and Fig.~\ref{fig:ics_grid}
shows the poorer accuracy of the resulting particle models.
Especially for the models with tangentially anisotropic initial conditions,
the smaller number of particles on radial orbits together with GWR
makes it more difficult to reproduce the radially anisotropic target.

We conclude that with the new regularization method
NMAGIC converges to the (theoretically essentially unique) solution
to a very good level of accuracy,
independently of the choice of the initial particle model. 
In this respect, the new MPR method is a definite improvement
over the traditional weight entropy scheme.

\section{Effects of imperfect data}
\label{ssec:real}
In astronomical applications, the data constraints
are typically less stringent than in the idealized case considered so far.
We therefore now investigate how the results change
in more realistic circumstances.

The following tests represent 
a sequence of problems that are increasingly less determined by the data,
starting from the truncated target covered by \textsl{realistic} data (Sec.~\ref{sssec:real}),
and moving on to infinite stellar systems 
constrained by data with finite extent. 
This allows us to isolate the different roles played by 
the quality and completeness of data,
the initial particle model, and the regularization scheme.

We find that it is still possible to get close to the target dynamical structure
from different initial particle models with the help of the new regularization method,
even though the lack and/or poor quality of data introduce degeneracies in the models.

\subsection{Truncated target and realistic kinematic errors}
\label{sssec:real}

First we construct NMAGIC models for a truncated target 
with the \textsl{realistic} kinematic data,
with the goal to establish how well the target galaxy can then be reproduced
from different initial particle systems.

The \textsl{realistic} data have larger error bars and smaller data coverage
(see Section~\ref{ssec:obs2}).
For these models $r_{\rm max}=100''$ is thus smaller
than in the previous case. The different initial particle models 
have the same anisotropy structure as in Section~\ref{sssec:ics}, 
but are adapted to this $r_{\rm max}$
- hence they are more similar to the target.

We have repeated the analysis described in Section~\ref{sssec:mu}
to determine the optimal value of the smoothing parameter $\mu$
when these \textsl{realistic} data are adopted.
Results do not change much, and suggest that we can keep
the values of $\mu=10^4$ for the GWR and $\mu=10^6$ for MPR.

\begin{figure}
\begin{center}
\subfloat{\includegraphics[angle=-90.0,width=8cm]{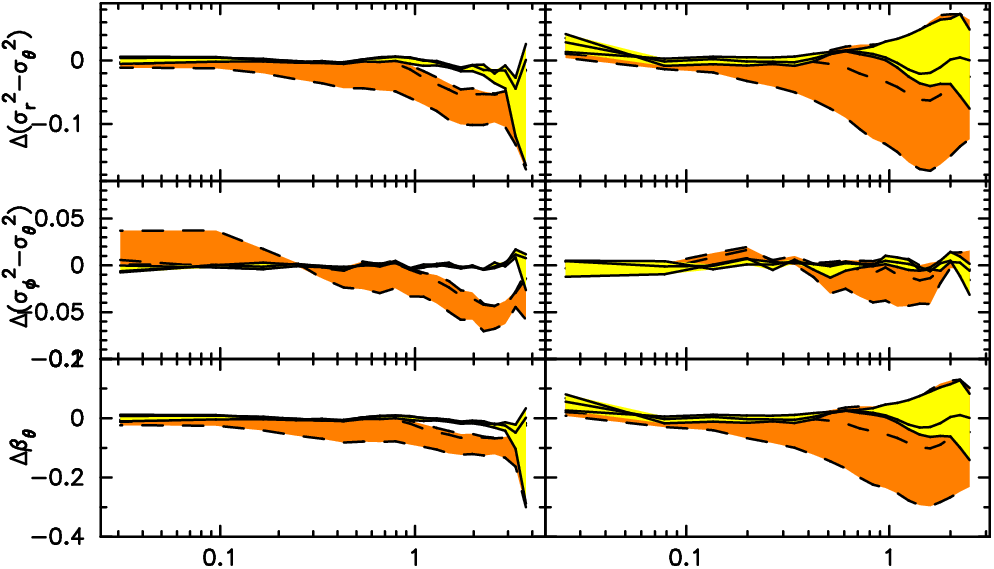}}\\
\subfloat{\includegraphics[angle=-90.0,width=8cm]{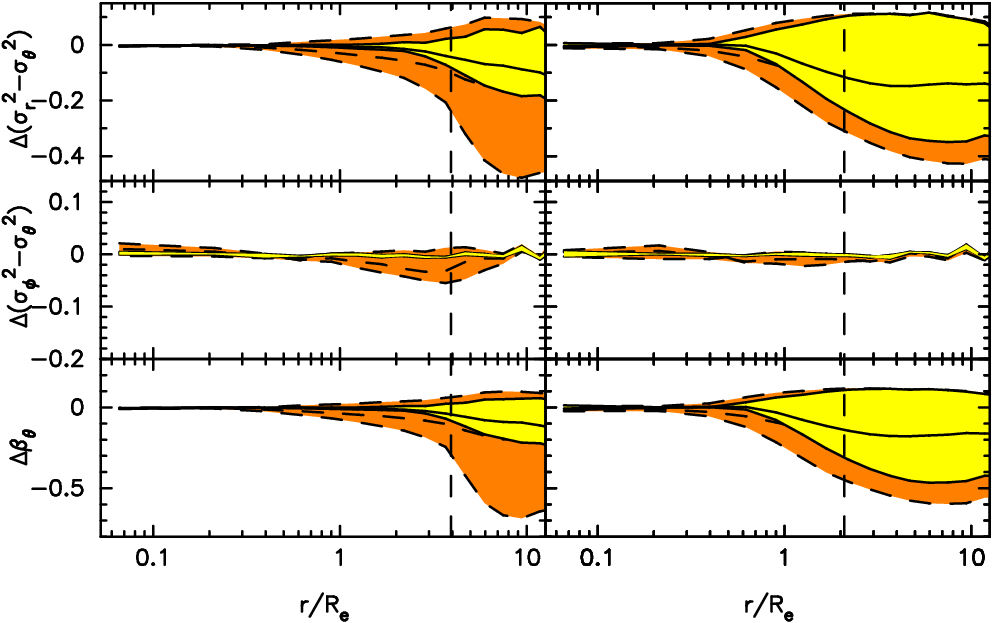}}
\caption[]{Recovery of the intrinsic kinematics for the truncated 
and infinite targets (top and bottom figure, respectively) 
with \textsl{idealized} and \textsl{realistic} data (left and right columns, respectively).
The vertical dashed line corresponds to the radial extent of the data
for the infinite target.
The shaded yellow (orange) area shows the range of deviations for the MPR (GWR) method,
when the specified range of initial conditions is adopted.
Full (dashed) lines represent the deviations for the final NMAGIC models
obtained with MPR (GWR) starting from different initial particle models.
Plotted in each panel are, from top to bottom, deviations 
of normalized $\sigma^2_r-\sigma^2_\vartheta$, $\sigma^2_\phi-\sigma^2_\vartheta$,
and anisotropy parameter $\beta$ from the respective true value of the target; 
see Sections~\ref{sssec:real} and~\ref{sssec:finite_data}.
}
\label{fig:delta_intkin}
\end{center}
\end{figure}

\begin{figure}
\includegraphics[angle=-90.0,scale=0.6]{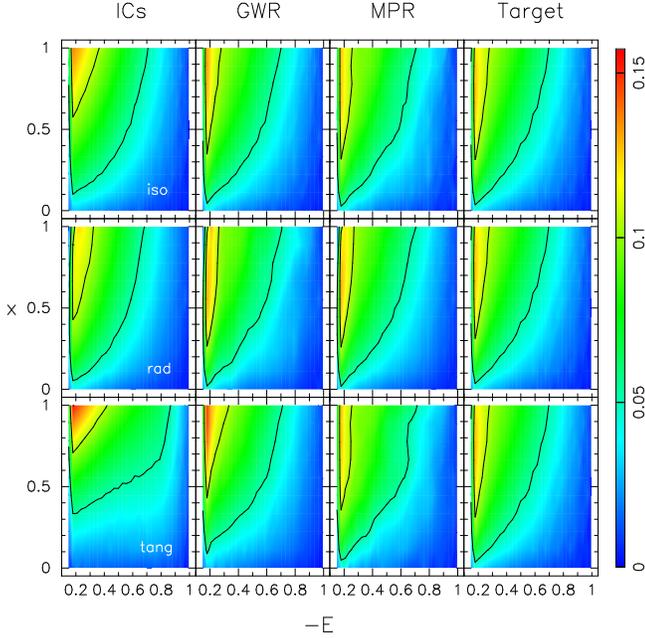}
\caption[]{Recovery of the MDF of the truncated target (last column on the right)
with \textsl{realistic} data for different initial particle models. As Fig.~\ref{fig:ics_grid}.
}
\label{fig:ics_grid_real}
\end{figure}

The results of these models are shown in Figs.~\ref{fig:delta_intkin} (top part)
and~\ref{fig:ics_grid_real}, and more quantitatively in Table~\ref{tab:trunc_ics},
for both regularization methods.
The top part of Fig.~\ref{fig:delta_intkin} shows the deviations 
of the models from the target,
comparing the two cases in which \textsl{idealized} and \textsl{realistic} constraints are used.
The three subpanels show the deviations of $\sigma^2_r-\sigma^2_\vartheta$
and $\sigma^2_\phi-\sigma^2_\vartheta$ normalized by the sum
of the two velocity dispersions, and the deviations of the velocity anisotropy $\beta$,
as a function of radius.
In this figure, the shaded regions correspond to the range of deviations 
obtained from modelling the data with the chosen initial particle systems.

As intuitively might be expected, these deviations increase $(i)$ moving to larger radii,
and $(ii)$ when \textsl{realistic} data are considered.
The effect of imperfect data on the final models
is noticeable, in particular closer to the model boundary
where poorer constraints from slit data worsen
the recovery of the intrinsic kinematics.

However, we see that also for realistic data
the new MPR works well in recovering the internal kinematics
of the target galaxy independently of the initial particle model,
and it is superior to the GWR, as deviations are considerably smaller.

The accuracy with which the phase-space mass distribution function of the target
is matched by the NMAGIC models is shown in Fig.~\ref{fig:ics_grid_real},
where is clear that the traditional GWR works less well with these realistic data, 
especially when tangential initial conditions are adopted.

Because the realistic error bars are larger than those used in the experiments in Section~\ref{sec:test}, 
the particle weights undergo smaller changes
until they match target observables in a $\chi^2$ sense
[see equations~(\ref{eqn:chi}) and~(\ref{eqn:delta})].
For the same reason, the final normalized $\chi^2$ 
between data and model observables turns out to be smaller
(see Table~\ref{tab:trunc_ics}).

To conclude, these experiments show that 
the new MPR method improves both the quality with which the intrinsic properties of the target galaxy
can be recovered, and the independence of the final particle model
from the adopted initial model.
In fact, MPR makes it possible to recover the underlying dynamical structure 
of our radially anisotropic target galaxy
with good accuracy ($\Delta\beta\sim\pm0.1$ only at the outermost point) 
even when the quality of the data is not perfect.
As already noticed, MPR allows the use of higher $\mu$ values,
thus reducing mass fluctuations and enforcing the smoothness of the underlying model
without spoiling the fit to the data.

\subsection{Finite data for an infinite target}
\label{sssec:finite_data}
Real stellar systems are clearly not as sharply truncated in radius 
as the target galaxies studied so far,
and their outer regions are usually not constrained by the available data.
We now come to the more realistic case of modelling an infinite target galaxy using finite data,
to explore the limitations that the modelling encounters in this case.

As target galaxy we consider our usual Hernquist sphere with scale length $a=1$ kpc,
but this time without truncation.
Because of the extreme behaviour of the OM radially anisotropic systems
at large radii, we choose a milder anisotropy for our target galaxy,
using the models of \citet{gerhard91} with specified circularity function 
[equations~(2.2) and~(3.14) therein],
which only depend on a constant parameter $\alpha$, set equal to 2.

Following the same procedure as adopted above,
we model the target starting from different initial particle models
(isotropic, less radially anisotropic, and more radially anisotropic
than the target) and using both regularization schemes.
\textsl{Idealized} data and \textsl{realistic} data are considered in turn.

\subsubsection{Infinite target with excellent but radially limited data}
Our modelling of the infinite target 
with \textsl{idealized} but radially limited data
confirms previous experiments \citep[][\citetalias{dl08}]{thomas04}
in that the velocity dispersions, streaming rotation, and anisotropy parameter
of the infinite target galaxy can be reproduced reliably
only in the regions well inside the part of the galaxy covered by the data.

Quantitative results are reported in Table~\ref{tab:trunc_inf},
while the bottom left panel of Fig.~\ref{fig:delta_intkin} 
shows the range of deviations of the final models from the target
obtained for different initial particle systems.
\begin{table}
\centering
\begin{tabular}{l l c l c c c}
\hline
&  & ICs & $\chi^2/J$  & $(\chi^2/J)_{\rm A_{lm}}$ & $(\chi^2/J)_{\rm slit}$\\ 
\hline
\textsl{Idealized}   & GWR  & iso         & 1.08  &  0.39 & 1.17 \\
data                 &      & $\alpha=1$  & 1.02  &  0.27 & 1.11 \\
                     &      & $\alpha=3$  & 0.89  &  0.21 & 0.98 \\
                     & MPR  & iso         & 0.92  &  0.45 & 0.98 \\
                     &      & $\alpha=1$  & 1.02  &  0.34 & 1.12 \\
                     &      & $\alpha=3$  & 0.78  &  0.34 & 0.86 \\
\hline
\textsl{Realistic}   & GWR  & iso         & 0.54 & 0.27 & 0.67 \\
data                 &      & $\alpha=1$  & 0.46 & 0.19 & 0.59 \\
                     &      & $\alpha=3$  & 0.43 & 0.16 & 0.56 \\
                     & MPR  & iso         & 0.62 & 0.27 & 0.79 \\
                     &      & $\alpha=1$  & 0.58 & 0.25 & 0.74 \\
                     &      & $\alpha=3$  & 0.55 & 0.22 & 0.71 \\
\hline

\end{tabular}
\caption{NMAGIC models for the infinite target galaxy in its fixed potential.
Different initial conditions and regularization schemes are adopted, 
as explained in Section~\ref{sssec:finite_data}. $\chi^2$ is the usual goodness of fit,
normalized by the respective number of observables. }
\label{tab:trunc_inf}
\end{table}
The vertical dashed line corresponds to the outermost data point.
This panel shows that in the inner regions of the galaxy, 
where the data provide good constraints to the models,
the intrinsic properties of the target galaxy 
are well recovered independently of the initial particle model,
as already found for the truncated target galaxy.
However, at larger radii, and close to the outermost data point,
regularization plays a dominant role in the weight correction of particles,
and in those external regions a bias towards the dynamical structure 
of the initial particle model cannot be avoided.

Nevertheless, our experiments show that 
the new MPR considerably reduces such bias towards
the dynamical structure of the initial particle model,
as can be seen comparing the two shaded regions
for MPR and GWR.

If we require $|\Delta\beta|\leq0.1$ and compute how far out this is achieved
for the range of models obtained starting from different initial conditions,
we find that this radius is $1.4R_{\rm eff}$ for GWR,
while it shifts to $4.3R_{\rm eff}$ when adopting MPR.
Considering instead the radius $r(\Delta\beta=\pm0.2)$, 
the standard GWR fails at $3.1R_{\rm eff}$, while the new method at $8 R_{\rm eff}$.

When, as in this case, a range of dynamical models obtained from
different initial particle models is compatible with the data, one
could compare and rank models according to the usual goodness-of-fit
basis (see \eg Table~\ref{tab:trunc_inf}), or additionally according to a
plausibility criterion that, \eg, favours a constant or smooth outer
anisotropy profile.

\subsubsection{Infinite target with realistic and finite data}

As a logical final step, we consider the case in which an infinite target
like the one described above is constrained by realistic, rather than idealized, data.

We model this target starting again from different initial particle models,
and show the accuracy of the final NMAGIC models in 
the bottom right panel of Fig.~\ref{fig:delta_intkin}, 
and in Table~\ref{tab:trunc_inf}.

The bottom part of Fig.~\ref{fig:delta_intkin} 
compares the deviations of the final models from the target
for \textsl{idealized} and \textsl{realistic} data.
Apparently, the realistic constraints
on an infinite target galaxy make it really hard
for NMAGIC to recover the true intrinsic kinematics of the target 
independently of the initial particle model,
even though it is still true that the new MPR method is superior to the GWR. 

To quantify how well the particle model reproduces
the intrinsic kinematics of the target galaxy, we can compute 
$r(\Delta\beta=\pm0.1)$, which  is $0.6R_{\rm eff}$ for the standard GWR,
and $1R_{\rm eff}$ when adopting the new regularization.
Considering instead the radius $r(\Delta\beta=\pm0.2)$, 
GWR fails at $0.9R_{\rm eff}$, while MPR at $1.4R_{\rm eff}$.
Here the kinematic data extend to $\sim2R_{\rm eff}$.

Thus, the results previously obtained for the \textsl{idealized} data are confirmed:
the new regularization scheme provides a better
reconstruction of the target properties,
and is more independent on the choice of the initial particle model.
However, as soon as there is a lack of data to constrain the models,
regularization becomes the dominant term in the force of change acting on particle weights,
and the bias towards the initial particle model becomes evident.

The main conclusion from these tests is that
the reliability of our dynamical models is limited
to those regions in which good observational data exist,
and that the better quality of the data is reflected
in a better recovery of the intrinsic properties of the target galaxy.

\section{Regularized particle models for NGC~4697 and NGC~3379}
\label{sec:gal}
\begin{figure*}
\centering
  \includegraphics[angle=-90.0,width=0.45\linewidth]{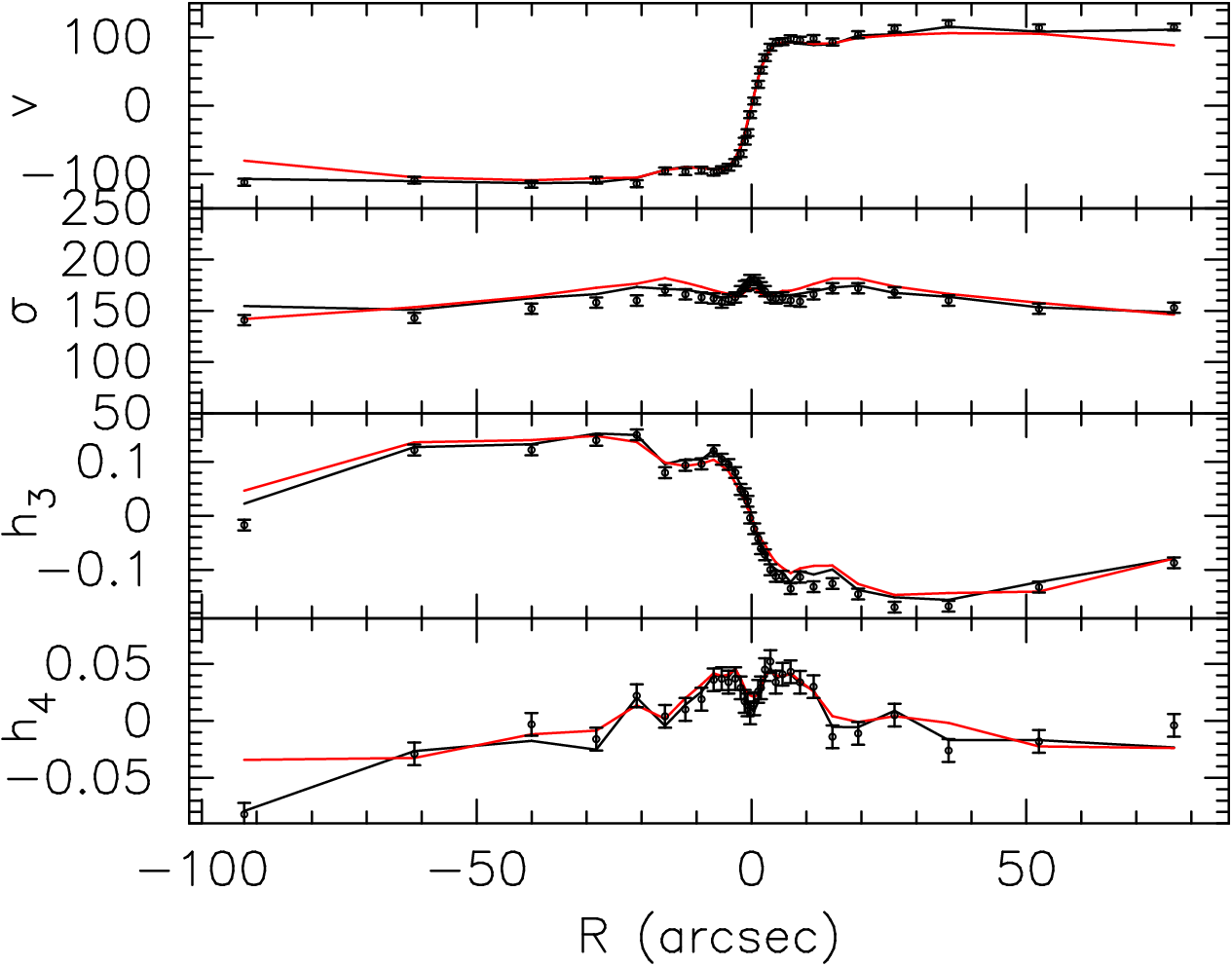}
  \includegraphics[angle=-90.0,width=0.45\linewidth]{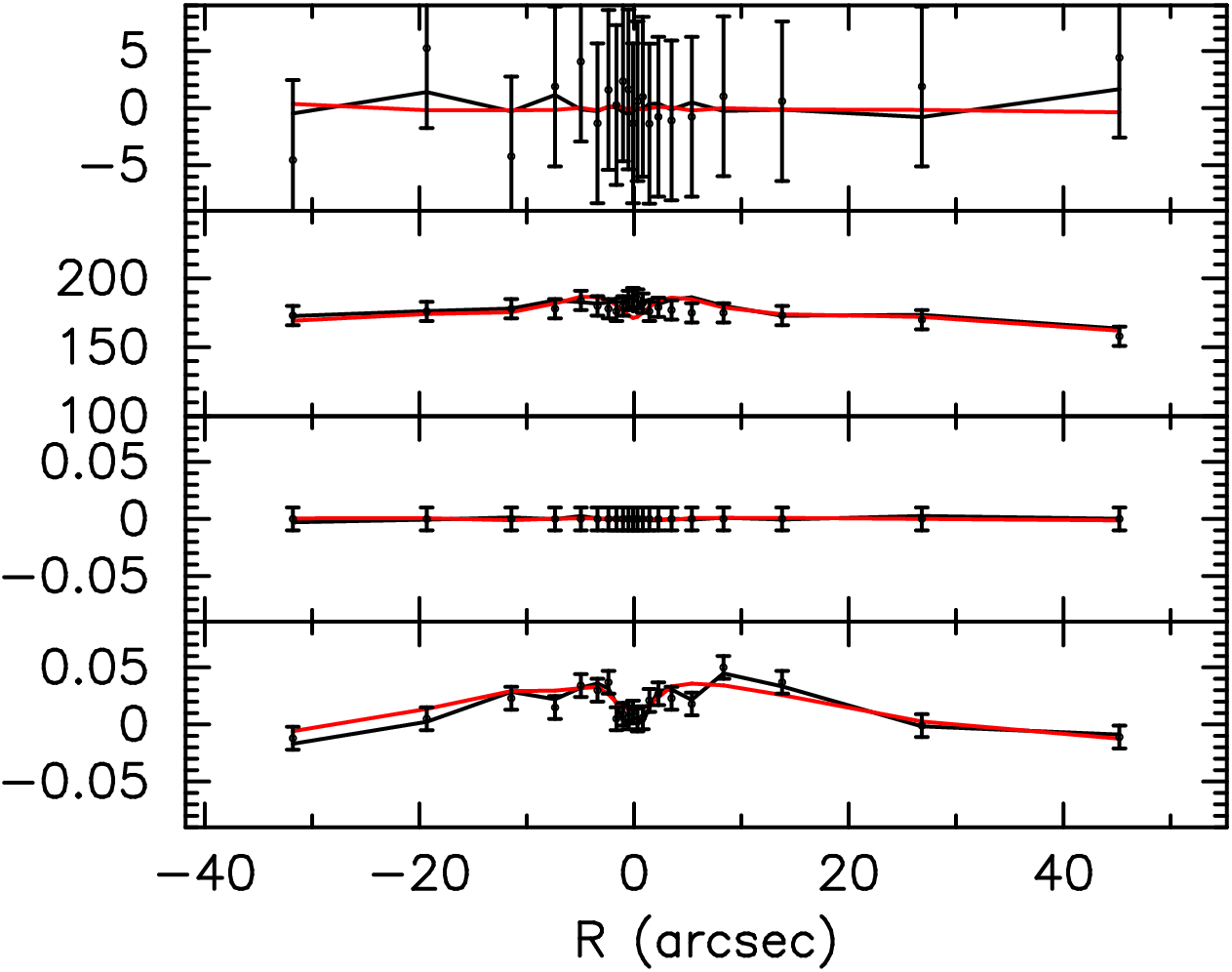}
\caption{Particle model fits to the slit data of NGC~4697 
along the major (\textit{left}) and minor axis (\textit{right}).
The model data points are averages over the same slit cells as the
target data, and are connected by
straight line segments. The \textit{black line} shows model J of \citetalias{dl08},
while the \textit{red line} shows the same model obtained using the new MPR.}
\label{fig:slits_j}
\end{figure*}
We now show our new regularization method
at work on two real galaxies,
and reconstruct the best-fitting NMAGIC models
determined in \citetalias{dl08} and \citetalias{dl09}
for the two intermediate luminosity elliptical galaxies
NGC~4697 and NGC~3379, respectively.

\citetalias{dl08} and \citetalias{dl09} used NMAGIC
to fit spherical and axisymmetric models of different inclinations
to extensive data sets for these galaxies, including photometry,
long-slit spectroscopy, integral-field data and PNe velocities.
Different from the experiments of Sections~\ref{sec:test} and~\ref{ssec:real},
particles are evolved in a total gravitational potential
\begin{equation}
\phi = \phi_{\star}+\phi_D,
\end{equation}
where $\phi_{\star}$ is estimated from the $N$-particle model for the light distribution
via a spherical harmonic decomposition \citep[\citetalias{dl07}]{sellwood03} 
assuming a constant mass-to-light ratio $\Upsilon$,
and $\phi_D$ is a dark matter halo potential with the logarithmic parametrization
\begin{equation} \label{eq:log_halo}
 \phi_D(R,z)=\frac{v_0^2}{2}\ln(r_0^2+R^2+\frac{z^2}{q^2}).
\end{equation}
Moreover, the mass-to-light ratio is not a fixed parameter,
but rather it is determined simultaneously with the modelling
of the dynamical structure in the NMAGIC run.

For both galaxies, the slit data show clear rotation
along the major axis on the sky.
However, the regularization scheme as developed in Section~\ref{sssec:loc_priors}
for spherical systems discourages any rotation in the particle model,
as it biases individual weights towards the same prior
regardless of the sense of rotation of the particles.
Thus, in the following tests we adopt a modified setup
for the grid of priors, binning particles according to
$E$ and $x$, and also according to the sign of their $L_z$, 
and assigning individual priors 
that differ between particles with different sense of rotation.
For an axisymmetric potential, this effectively uses
the total angular momentum $L$ as an approximate third integral,
which may be expected to be a reasonable first approximation
unless $L\simeq L_z\simeq0$ \citep{gersaha91}.

\subsection{The case of NGC~4697 and its halo}
The intermediate luminosity elliptical galaxy NGC~4697
is seen almost edge-on.
Assuming that the observed nuclear dust-lane is settled in the equatorial plane,
\citet{dejonghe96} derived an inclination $i=78\degrees\pm5\degrees$,
which is consistent with the bulge-disk decomposition of \citet{scorza98} 
if the disk component has an intrinsic axis ratio $h/R\sim0.2$.

NGC~4697 has fitted Sersic model index $n=3.5$,
and an effective radius $R_{\rm eff}\approx 66''=3.36$ kpc
at an assumed distance $D=10.5$ Mpc.
Kinematic data show significant major axis rotation reaching $\sim100$ km/s at $90''$.

NMAGIC axisymmetric particle models assuming an inclination $i=80\degrees$
were constructed for NGC~4697 (\citetalias{dl08})
fitting simultaneously surface brightness photometry,
long-slit absorption-line kinematics
and hundreds of PNe velocities.
A range of quasi-isothermal halos was found to be consistent 
with the observational constraints, 
and a massive halo with circular velocity $v_0=250$ km/s at $4.3R_{\rm eff}$,
referred to as model J in the notation of \citetalias{dl08},
fits the PN data best.
This model is characterized by a moderately radially anisotropic orbit distribution,
with the anisotropy parameter $\beta\sim0.3$ at the center 
and increasingly higher in the outer regions.

These models were constructed using the traditional GWR, 
and the $\mu$ parameter was set to $100$ 
to avoid strong biases to the initial conditions.
This in turn led to some overfitting of the slit kinematics data,
especially for the higher order kinematic moments (see \citetalias{dl08} for details). 

Thus we now build a new regularized J model of NGC~4697,
to see whether a similarly good but smoother particle model
can be obtained with the help of the new MPR.
We rerun that exact model with the code NMAGIC
using the new regularization scheme, and $\mu=10^5$,
and using constraints from both photometric and kinematic data, including PNe.
As specified above, we bin particles according to their integrals $E,x$,
and $L_z$ when adopting the new regularization method,
to allow for the rotation seen in the slit data.

\begin{figure}
\includegraphics[angle=-90.0,width=1.\hsize]{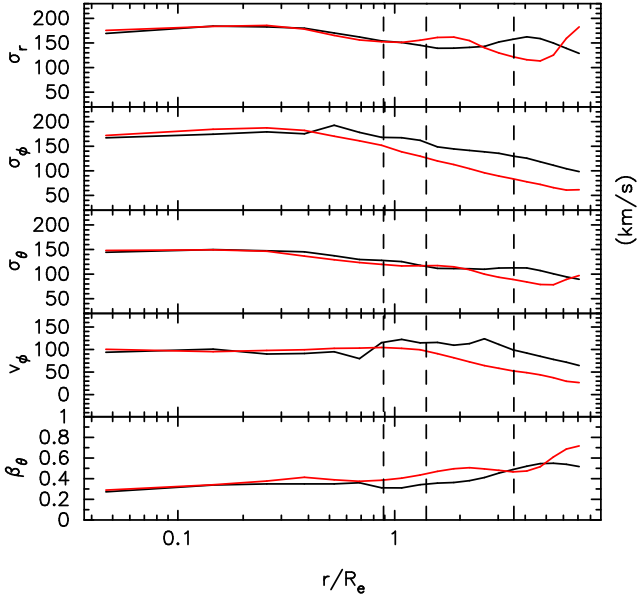}
\caption[]{Internal velocity moments in the equatorial plane for
model J of \citetalias{dl08} (\textit{black line}) and the new regularized J model
obtained here (\textit{red line)}, for NGC~4697. 
The vertical dashed lines indicate the radial extent of the minor axis slit data, 
major axis slit data, and PN data, from left to right.}
\label{fig:int_j}
\end{figure}

A comparison of the final particle models
obtained by \citetalias{dl08} and with this new MPR
is shown in Figs.~\ref{fig:slits_j} and~\ref{fig:int_j}.
Fig.~\ref{fig:slits_j} shows the projected absorption line kinematics
of the final particle models overplotted on the data points.
As discussed in \citetalias{dl08}, 
asymmetries between the left-hand side and right-hand side in the profiles 
do not imply deviations from axisymmetry or equilibrium,
but rather they are due to averages over slightly different slit cells on both sides.
As expected, these asymmetries decrease when using the new MPR,
which allows a higher amount of regularization,
and the model profiles are indeed smoother than for the \citetalias{dl08} model.

The intrinsic kinematics of the final particle models
are compared in Fig~\ref{fig:int_j}:
the velocity anisotropy increases from the center outwards
when adopting either GWR or MPR, but MPR results in
much smoother profiles in the regions constrained by data.
\subsection{The case of NGC~3379 and its halo}
In \citetalias{dl09} a sequence of spherical and axisymmetric NMAGIC models 
fitting an extensive data set 
(photometry, long-slit and SAURON absorption-line kinematics, 
PN velocity dispersion data) was constructed to investigate 
the mass distribution and orbital structure 
of the intermediate luminosity E1 galaxy NGC~3379.

No independent information on the inclination is available for this galaxy,
and values of $i>40\degrees$ are consistent with the photometry.
The effective radius $R_{\rm eff}\approx 47''=2.23$ kpc
at an assumed distance $D=9.8$ Mpc.
The combined kinematic data set shows major axis rotation reaching $\sim50$ km/s
at $20''$, with PNe indicating a further increase to $\sim70$ km/s at $220''$.

\citetalias{dl09} explored a sequence of spherical and axisymmetric models,
together with some triaxial test models.
They found that their results were insensitive to the adopted geometry.
Both strongly radially anisotropic models 
embedded in massive dark matter halos and nearly isotropic systems dominated by the stellar mass
are consistent with the data.
I.e., even the extensive data set used in the modelling
was not sufficient to break the mass-anisotropy degeneracy \citep{binmam82}
because of the rapidly decreasing velocity dispersion profile for NGC~3379.
However, an analysis of the quality of the fit
and of the likelihood of the observed PN velocity data
for the spherical models
slightly favoured a range of models centered around 
the radially anisotropic halo C,
which was obtained for a quasi-isothermal potential 
in equation~(\ref{eq:log_halo}),
with $r_0/R_{\rm eff}=3$, $v_0=130$ km/s, and $q=1$.

We now reconstruct that spherical C model with the new MPR method.
Given the data setup is very similar
to the one adopted in the tests of Section~\ref{ssec:real},
we set $\mu=10^6$.
We bin particles according to their integrals $E,x$,
and $L_z$, because of the observed rotation in both the slit and SAURON data.
For the energy grid we use a linear binning of the function $\exp(E)$, 
which provides a better sampling of the model DF in the outer regions for this potential.

\begin{figure}
\centering
\includegraphics[angle=-90.0,scale=.55]{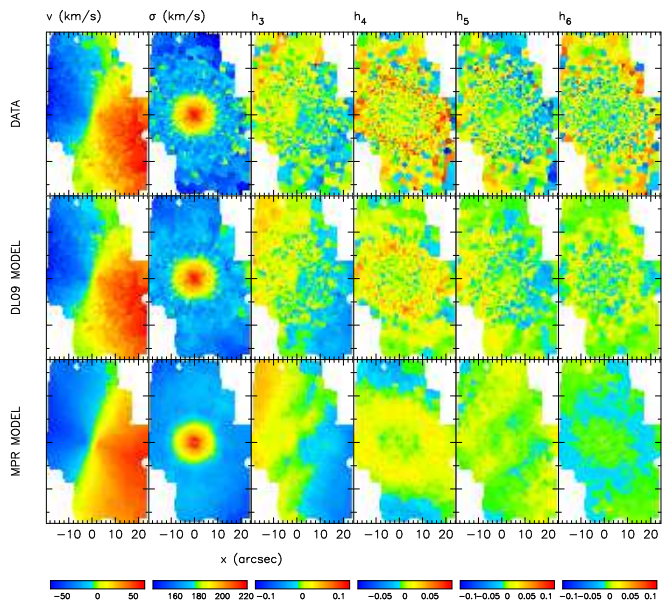}
\caption[]{Particle model fits to the SAURON integral field kinematic data for NGC~3379.
Top row: symmetrized SAURON data. Middle row: best-fitting C model (\citetalias{dl09}).
Bottom row: new regularized C model. Mean velocity, velocity dispersion, 
and higher order Gauss-Hermite moments are shown in the panels from left to right.}
\label{fig:sauron_c}
\end{figure}

Fig.~\ref{fig:sauron_c} shows the fit to SAURON data
for the final NMAGIC models obtained with GWR and MPR.
Both particle models reproduce the observed rotation with great accuracy,
and the MPR model is clearly smoother than the original (symmetrized) data.
In particular, notice the ring-like structure
in the $h_4$ plot. 
Even though the new model is generated using a much higher value of $\mu$,
i.e. much stronger smoothing, it still does a good job in fitting the observational data,
with $(\chi^2/J)_{\rm sauron}=0.86$ (compared to 0.17 for the traditional GWR).

\begin{figure}
\includegraphics[angle=-90.0,width=1.\hsize]{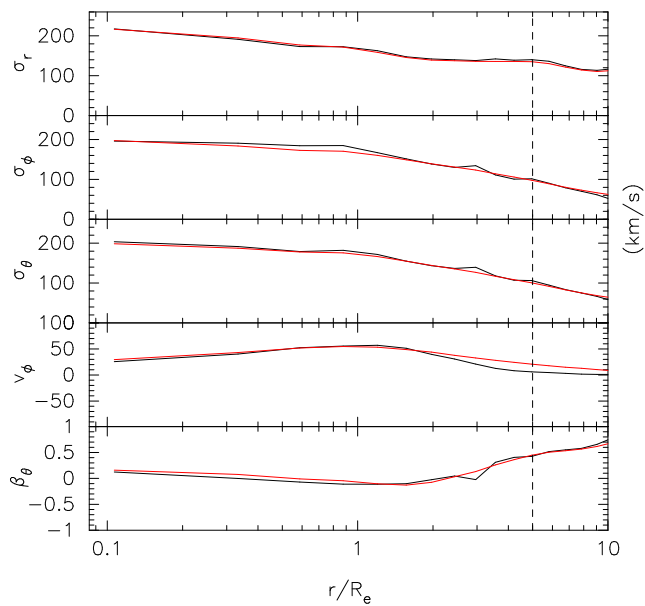}
\caption[]{Internal velocity moments in the equatorial plane for
model C of \citetalias{dl09} (\textit{black}) and the new regularized C model
obtained here (\textit{red)}, for NGC~3379. The vertical dashed line marks the last data point.}
\label{fig:int_c}
\end{figure}

The intrinsic kinematics of the final NMAGIC models (Fig.~\ref{fig:int_c}) are similar,
but using MPR the kinks in the profiles disappear.
It can be seen that a strong radial anisotropy is required 
to match the PNe data in this dark matter halo.

\section{Discussion and conclusions}
\label{sec:discuss}
Building on the work of \citetalias{st96}, 
successive investigations \citep[\citetalias{dl07}; \citetalias{dl08}; \citetalias{dl09};][]{dehnen09,longmao10}
have shown the power of the $\chi^2$M2M modelling technique
to learn about the dynamics of galaxies.
$\chi^2$M2M methods work by adapting the weights of an $N$-body particle system 
until the observational data are well matched in a $\chi^2$ sense, 
subject to additional regularization constraints.
These constraints are needed to prevent the particle model
from acquiring large fluctuations 
because of scatter and noise especially in the kinematic data.

Traditionally, a Global Weight entropy Regularization (GWR) is adopted
to regularize the underlying particle system.
However, through constant flat priors GWR introduces a bias in the particle model
which makes it difficult to reproduce strong phase-space gradients
of the target galaxy, \eg anisotropic velocity distributions,
unless its dynamical structure is known beforehand. 

In this paper we have described a new Moving Prior Regularization (MPR) method,
based on a prior distribution for the particles which evolves with the model.
Individual particle priors are updated along with particle weights
to keep track of the phase-space structures of the evolving weight distribution.
The basic idea is to determine the priors such that they are similar for
particles on neighbouring 
orbits, specified by orbital invariants or integrals of motion such as
energy and angular momentum in the spherical case. 
The new priors are then used in a weight entropy function to ensure a regularization
which smoothes locally in phase-space, without erasing global phase-space gradients.

We have then tested this MPR scheme,
together with the $\chi^2$M2M modelling technique,
using a series of spherical target galaxies
with both idealized and realistic data.
Our main conclusions are as follows:
\begin{itemize}
\item For a truncated spherical target galaxy with idealized data,
for which in theory a unique inversion of the data exists,
our NMAGIC models with MPR show that the target can be recovered 
accurately, and independent of the initial particle model. 
\item The new MPR generally improves both the accuracy 
with which the dynamical structure of the target galaxy is reproduced,
and the convergence to the true solution independent of the initial particle model.
Compared to GWR, biases in the anisotropy structure are removed,
and local fluctuations in the intrinsic distribution function are reduced.
Moreover, MPR allows a higher amount of smoothing
than the weight entropy regularization, while the data are still fitted well.
\item Lack or poorer quality of data introduce degeneracies 
in the dynamical modelling results and a dependence on the initial particle
model,
so that the reliability of the models is limited
to those regions in which good observational data exist.
Also in this case, the new MPR achieves a better
reconstruction of the target properties
and is less dependent on the choice of the initial particle model.
\item Using the new MPR, we have reconstructed the best-fitting NMAGIC models
determined in previous work by \citetalias{dl08} and \citetalias{dl09}
for the two elliptical galaxies NGC~4697 and NGC~3379 in their dark matter halos.
To this goal, we have extended the MPR method 
to the axisymmetric case, using the integrals $E$ and $L_z$ and the 
total angular momentum as an approximation to the third integral.
The final models are intrinsically smoother 
and provide smoother fits to the available data.
\end{itemize}

There is clearly room for improving the current version of MPR:
the method could be generalized to systems of lower symmetry 
using the invariants associated with orbits, \eg the turning points,
to assign moving priors in phase-space to the particles.
Moreover, a cumulative grid-less variant of the method
could also be implemented. Re-sampling of the $N$-body system
from time to time during and after the adjustment of the weights \citep{dehnen09}
would enforce equal weight for particles
orbiting the same torus, but it would not take care 
of smoothing between nearby tori with very different weights.

To conclude, the experiments described in this paper
show that the moving prior regularization method 
improves the correct and unbiased recovery
of the orbit structure of the target galaxy
from noisy data.
A similar regularization scheme could also be implemented
in Schwarzschild orbit superposition models.
\bigskip
\noindent

\section*{Acknowledgments}
We thank Karl Gebhardt for making 
his smoothing spline code available,
and an anonymous referee for a careful reading of the paper.
LM acknowledges support from and participation in 
the International Max-Planck Research School on Astrophysics 
at the Ludwig-Maximilians University.
 
\bibliographystyle{mn2e}
\bibliography{nmagic}

\label{lastpage}

\end{document}